\documentclass[a4paper,USenglish]{lipics-v2019}
\pdfoutput=1

\usepackage{xspace}
\usepackage[algo2e,boxed,vlined,linesnumbered]{algorithm2e}

\usepackage{placeins}

\usepackage{tabularx}

\newcommand{\Trule}{\rule{0pt}{3ex}}
\newcommand{\Brule}{\rule[-1.5ex]{0pt}{0pt}}

\newcommand{\optprobA}[3]{
\begin{center}
\begin{tabularx}{\columnwidth}{|l X|}
	\hline
	\multicolumn{2}{|c|}{#1\Trule} \\
	\textbf{Input:\ }&{#2}\\
	\textbf{Solution:\ }&{#3\Brule}\\
	\hline
\end{tabularx}
\end{center}
}
\usepackage{etoolbox}

\newcommand{\sv}[1]{}
\newcommand{\lv}[1]{#1}

\newcommand{\appendixText}{}

\newcommand{\toappendix}[1]{#1}

\DeclareMathOperator{\poly}{\operatorname{poly}}
\DeclareMathOperator{\OPT}{\operatorname{OPT}}
\DeclareMathOperator{\mc}{\operatorname{mc}}
\DeclareMathOperator{\dist}{\operatorname{dist}}
\DeclareMathOperator{\starRatio}{\operatorname{r}}
\DeclareMathOperator{\starRatioCur}{\operatorname{r^{cur}}}

\let\star\relax 
\DeclareMathOperator{\star}{\operatorname{st}}

\SetKwProg{Pn}{Function}{:}{\KwRet}
\def\assign{\leftarrow}
\SetKw{Continue}{continue}
\DontPrintSemicolon 

\renewcommand{\S}{Section}

\title{Approximation Algorithms for Steiner Tree Based on Star Contractions: A~Unified View}

\titlerunning{Star Contractions for Steiner Tree}

\author{Radek Hušek}{Computer Science Institute of Charles University, Charles University, Czech Republic}{husek@iuuk.mff.cuni.cz}{}{}

\author{Dušan Knop}{Department of Theoretical Computer Science, Faculty of Information Technology, Czech Technical University in Prague, Czech Republic}{dusan.knop@fit.cvut.cz}{https://orcid.org/0000-0003-2588-5709}{Supported by the OP VVV MEYS funded project CZ.02.1.01/0.0/0.0/16 019/0000765 ``Research Center for Informatics''.}

\author{Tomáš Masařík}{\lv{Faculty of Mathematics, Informatics and Mechanics, }University of Warsaw, Poland \sv{\&}\lv{\\} \lv{Department of Applied Mathematics, }Charles University, Czech Republic}{masarik@kam.mff.cuni.cz}{https://orcid.org/0000-0001-8524-4036}{have received funding from the European Research Council under the European Union’s Horizon 2020 research and innovation programme Grant Agreement 714704.\\
\includegraphics[height=30px]{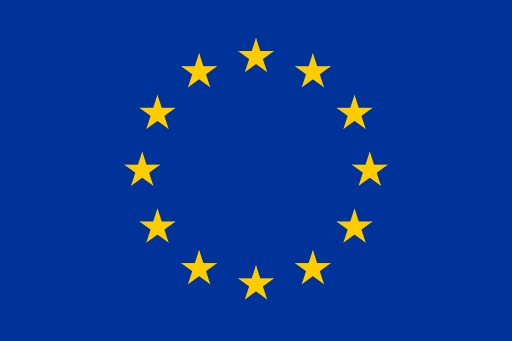}\quad\includegraphics[height=30px]{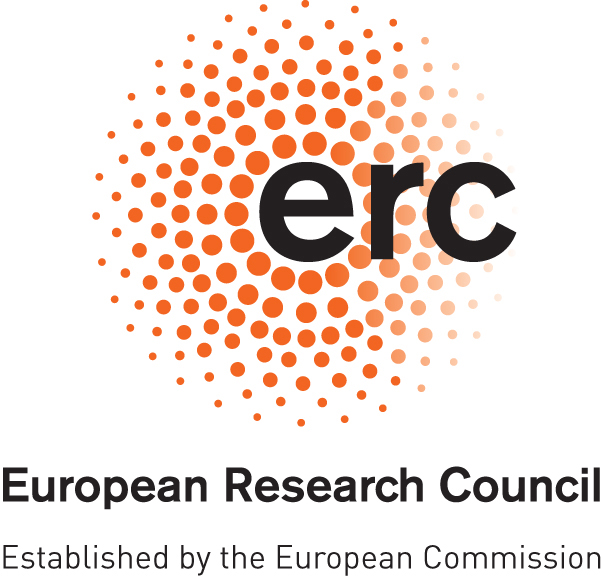}
}
\authorrunning{R.\ Hušek, D.\ Knop, and T.\ Masařík}

\Copyright{Radek Hušek, Dušan Knop, and Tomáš Masařík}



\category{}

\relatedversion{}

\supplement{https://github.com/JohnNobody-3af744f30980b7458372/star-contractions}

\lv{%
\funding{%
Students were supported by Charles University student grant SVV-2017-260452 and GAUK 1514217.
}}

\acknowledgements{}

\EventEditors{John Q. Open and Joan R. Access}
\EventNoEds{2}
\EventLongTitle{42nd Conference on Very Important Topics (CVIT 2016)}
\EventShortTitle{CVIT 2016}
\EventAcronym{CVIT}
\EventYear{2016}
\EventDate{December 24--27, 2016}
\EventLocation{Little Whinging, United Kingdom}
\EventLogo{}
\SeriesVolume{42}
\ArticleNo{23}
\nolinenumbers 
\hideLIPIcs  

\begin{document}
\let\paragraph\subparagraph
\maketitle

\begin{abstract}
  In the \textsc{Steiner Tree} problem we are given an edge weighted undirected graph $G = (V,E)$ and a set of terminals $R \subseteq V$.
  The task is to find a connected subgraph of $G$ containing $R$ and minimizing the sum of weights of its edges.
  \textsc{Steiner Tree} is well known to be \textsf{NP}-complete and is undoubtedly one of the most studied problems in (applied) computer science.
  \textsc{Steiner Tree} has been analyzed within many theoretical frameworks as well as from the practical perspective.

  We observe that many approximation algorithms for \textsc{Steiner Tree}
  \begin{enumerate}
    \item follow a similar scheme (meta-algorithm) and
    \item perform (exhaustively) a similar routine which we call \emph{star contraction}.
  \end{enumerate}
  Here, by a star contraction, we mean finding a star-like subgraph in (the metric closure of) the input graph minimizing the ratio of its weight to the number of contained terminals minus one.
  It is not hard to see that the well-known MST-approximation seeks the best star to contract among those containing two terminals only.
  Zelikovsky's approximation algorithm follows a similar workflow, finding the best star among those containing three terminals.

  We perform an empirical study of star contractions with the relaxed condition on the number of terminals in each star contraction.
  Our experiments suggest the following:
  \begin{itemize}
    \item if the algorithm performs star contractions exhaustively, the quality of the solution is usually slightly better than Zelikovsky's 11/6-approximation algorithm,
    \item on average the quality of the solution returned by the MST-approximation algorithm improves with every star contraction,
    \item the same holds for iterated MST (MST+), which outperforms MST in every measurement while keeping very fast running time (on average $\sim 3\times$ slower than MST),
    \item on average the quality of the solution obtained by exhaustively performing star contraction is about 16\% better than the initial MST-approximation, and
    \item we propose a more precise way to find the so-called improved stars which yield a slightly better solution within a comparable running time (on average $\sim 3\times$ slower).
  \end{itemize}
  Furthermore, we propose two improvements of Zelikovsky's 11/6-approximation algorithm and we empirically confirm that the quality of the solution returned by any of these is better than the one returned by the former algorithm.
  However, such an improvement is exchanged for a slower running time (up to a multiplicative factor of the number of terminals).
\end{abstract}

\section{Introduction}\label{sec:intro}
In the \textsc{Steiner Tree} problem an edge weighted graph $G = (V,E)$ is given together with the set of \emph{terminal} vertices $R \subseteq V$; the non-terminal vertices are called \emph{Steiner vertices}.
The task is to find a connected subgraph of $G$ containing all terminals and minimizing the sum of weights of its edges.
\textsc{Steiner Tree} was among the first problems shown to be \textsf{NP}-complete~\cite{MR51:14644} and is one of the most studied problems in computer science since then.
\optprobA{\textsc{Steiner Tree}}
{A graph $G = (V, E)$, a set of terminals $R \subseteq V$, and a weight function \mbox{$w\colon E \to \mathbb{N}$}.}
{A Steiner tree $F \subseteq G$ containing a path between any two terminals $s,t \in R$.}
\textsc{Steiner Tree} is important not only as an interesting graph-theoretic problem but it has many real-world applications e.g.\ in network design or VLSI design~\cite{hwang1992steiner}.
Thus, it is extensively studied by both theoreticians and practitioners.
There are many theoretical results studying (approximation) algorithms for \textsc{Steiner Tree}; for an overview see e.g.\ a survey~\cite{hwang1992steiner}.
We now discuss a few theoretical results important for our work.

\subsection{Star-Contraction and Approximation Algorithms}
The most basic approximation algorithm for \textsc{Steiner Tree} is the algorithm based on finding a minimum spanning tree (MST) in the metric closure of the input graph~\cite{KouMB81}.
This algorithm was later improved by Zelikovsky~\cite{Zelikovsky93} who used the finding of augmenting stars containing three terminals to improve the algorithm of Kou et al.~\cite{KouMB81} and was the first to beat the barrier of $2$ for the approximation ratio.
Here an augmenting star consists of a Steiner vertex and exactly three terminals such that if we contract the just defined star into a terminal and compute the weight of an MST, the weight of the thus obtained MST together with the weight of a contracted star is strictly smaller than the weight of the former MST.
This approach was later improved by Borchers and Du~\cite{BorchersD97}.
Furthermore, the current best theoretical approximation algorithm of Byrka et al.~\cite{DBLP:journals/jacm/ByrkaGRS13} with approximation ratio $\ln(4) + \varepsilon$ is in fact based on star contractions.

We observe that the above mentioned algorithms not only use star contractions\lv{\footnote{In fact, some of these algorithms use the so-called \emph{full-components}. A full component is a tree in the input graph where terminals may only be in leaves, that is, cannot be the inner vertices of such a tree.}} as the main tool but, on top of this, most of these algorithms follow a very similar meta-algorithm---see Algorithm~\ref{alg:highLevelOverview} and Example~\ref{ex:MSTHighLevelOverview} below.
\begin{algorithm2e}[bth]
\SetKwFunction{BestStar}{find\_best\_star}
\SetKwFunction{Contract}{contract}
\SetKwFunction{FindSolution}{find\_solution}
\SetKwFunction{Eval}{eval}
\SetKwFunction{Finish}{finish}

\KwIn{$G = (V,E)$, set of terminals $R$, parameters $k, \tau \in \mathbb{N}$, and functions \Contract{}, \Eval{}, \Finish{}}
\KwOut{A Steiner tree}

$G' \assign G, \ R' \assign R$ \;
\While{$|R'| > \tau$}{
  $C \assign $ \BestStar{$G'$, $R'$, $k$, \Eval} \;
  \If{\Eval{$C$} $< \infty$}{
    $G',R',S' \assign$ \Contract{$G'$, $R'$, $C$} \;
    $\mathcal{S} \assign \mathcal{S} \cup \{ S' \}$ \;
  }
  \lElse{
    \textbf{break}
  }
}
  \Return{ \Finish{$G$, $\mathcal{S}$, $R$} } \;

\caption{\label{alg:highLevelOverview}%
A unifying high-level framework of selected approximation algorithms  for \textsc{Steiner Tree}.
The \protect\BestStar{} function finds a star $C$ with at most $k$ terminals which among all such stars achieves the lowest value under the evaluation function \protect\Eval{}.
The \protect\Contract{} function (usually) contracts the star~$C$ (note that~$C$ and~$S'$ could possibly be different {e.g.}~$C$ can be in the metric closure of~$G'$).
}
\end{algorithm2e}
There, we argue that the simplest algorithm we consider, the MST-approximation, can be described in the framework given in Algorithm~\ref{alg:highLevelOverview}.
We justify the same for (our modification of) Zelikovsky's algorithm in \S~\ref{sec:algorithmDetails}, Example~\ref{ex:zelikovsky}.

\begin{example}[MST]\label{ex:MSTHighLevelOverview}
  Observe that in order to find a spanning tree of minimal weight it suffices to find the best star with two terminals, that is, the cheapest edge in the metric closure of the given graph.
  Furthermore, if we then contract such a path, we reduce the size of the terminal set by one.
  Thus, one can set the parameter $k = 2$ and $\tau = 1$ (as we perform star contractions exhaustively).
  The \Eval{} function gives the length of the shortest path (i.e., the total weight of the proposed $2$-star), the \Contract{} function contracts, and the \Finish{} function contracts the given collection of $2$-stars.
\end{example}

\paragraph{Unbounded Size of a Best Star.}
A recent result of Dvo{\v{r}}{\'{a}}k et al.~\cite{DvorakFKMTV18}, which proposes a novel algorithm in the framework of parameterized approximations, also falls in the framework suggested in Algorithm~\ref{alg:highLevelOverview}.
Surprisingly, their algorithm uses an unbounded value of the parameter $k$, the terminal-size of a best star.
Their algorithm, given a parameter $p$ and the desired approximation ratio $\varepsilon > 0$, runs in time $f(p, \varepsilon)\poly(|G|)$ and outputs a solution of cost at most $(1+\varepsilon) \cdot \OPT(p)$, where $\OPT(p)$ is the value of an optimal solution that uses at most $p$ Steiner vertices.
Let us now discuss in more detail why the algorithm of Dvo{\v{r}}{\'{a}}k et al.\ follows the proposed framework; we discuss further technical details later (see \S~\ref{sec:algorithmDetails}).
First we set the parameters $k = \infty$ and $\tau = c \cdot \frac{p^2}{\varepsilon^4}$ for a suitable constant $c$.
\begin{description}
  \item[\Eval{}]
  Let $C$ be a connected subgraph of $G'$, let $w(C)$ be the total weight of edges in~$C$, and let $R_C \subseteq R'$ be the set of terminals contained in~$C$.
  The function \Eval{$C$} returns the value \( \frac{w(C)}{|R_C| - 1} \,. \)
  \item[\BestStar{}]
  Since the parameter $k = \infty$, the function returns a connected subgraph $C$ minimizing the value \( \frac{w(C)}{|R_C| - 1}\) among all connected subgraphs with at least two terminals.
  \item[\Finish{}]
  We first contract all subgraphs $C$ obtained so far (i.e., we construct the graph~$G'$).
  Then the algorithm of Fuchs et al.~\cite{FuchsKMRRW07} is invoked on $G'$.
\end{description}
It is worth noting that the algorithm of Fuchs et al.\ computes an optimal solution in time $f(\tau) \cdot \poly(|G'|)$.
Note that a similar running time has been achieved already by Dreyfus and Wagner~\cite{DBLP:journals/networks/DreyfusW71}.

We conclude that ``contracting (best) star'' (full components) is a popular and successful technique in the design of approximation algorithms for \textsc{Steiner Tree} (i.e., in theory).
Note that if we take into account stars containing more terminals (i.e., we increase $k$), the approximation ratio of the returned solution does not increase.
Thus, natural questions arise:
\begin{enumerate}
  \item Do the star contractions behave well in practice?
  That is, is it possible to improve the total weight of a solution returned by well-studied heuristics ({e.g.}\ MST approximation) significantly when we first perform few rounds of best star contractions?
  \item Is it common to find large (nearly) best stars?
  That is, is there a significant fraction of all contracted stars containing more than {e.g.}~5 terminals?
  \item Is there any point of Pareto optimality, i.e., is it (for example) possible to reduce the number of terminals by 20\% only using only 10\% of total work while improving the solution by 80\%?
  \item Is there a way to improve classical approximation algorithms using the perspective of the best star contractions?
\end{enumerate}

\subsection{Our contribution}
In order to answer our questions we have implemented the algorithm for \textsc{Steiner Tree} of Dvo{\v{r}}{\'{a}}k et al.~\cite{DvorakFKMTV18} (with some modifications in order to be able to run some experiments).
It is worth noting that even though our work is experimental the core of our work is mostly proof-of-concept, that is, we experimentally investigate and evaluate quantitative features of star contractions.
It is possible to improve the quality of the solution returned by the MST approximation for \textsc{Steiner Tree} quite significantly if we allow the use of stars containing more terminals.
The improvement is by more than 12\% on average and it is worth pointing out that the number outliers is reduced significantly as well (see Figure~\ref{fig:aggregatedMST}).
On the other hand, if our routine is applied with more sophisticated approximation algorithms (our modifications of Zelikovsky's algorithm), then these improvements seem to be rather negligible.

We answer the second question in negative:
Vast majority of the contractions performed by Algorithm~\ref{alg:highLevelOverview} on our instances (see below for the descriptrion of the dataset used) are those of stars containing two or three terminals (see also bellow Table~\ref{tab:starSizes} and Figure~\ref{fig:starSizes}\sv{ in the appendix}).
Next, we propose an improved routine for finding best starts possibly containing more terminals (thus achieving better star ratio; see \S~\ref{sec:findingBestStar}).
We can see that with this improved routine the number of best stars containing more than~5 terminals improves from (roughly) 2\% to 5\% (on the selected dataset).
Despite this improvement, overall performance is not much better than with normal star contractions, however, the time consumption stays low (approximately three times as much, see Figure~\ref{fig:work}).
It is worth noting that for both routines the best star containing only two terminals is found (on average) in more than 75\% of all contractions performed during the execution of the algorithm.

Despite our former believes that were supported by the results of Dvořák et al.~\cite{DvorakFKMTV18}, the answer for the third question seems to be also negative.
Figure~\ref{fig:quality-work}\sv{ in the appendix} indicate that there is no threshold point for neither classical nor improved stars.

\begin{figure*}[h!]
  \begin{center}
  \includegraphics[width=0.8\textwidth]{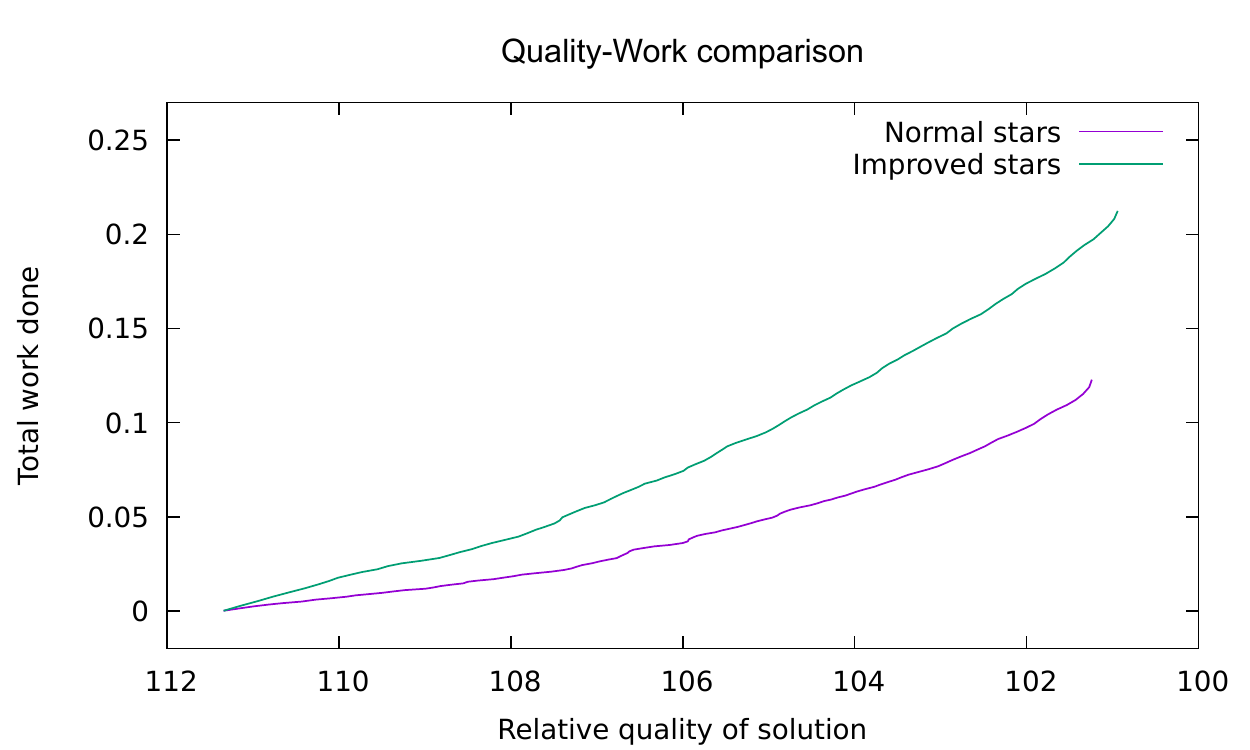}
\end{center}
\medskip
  \caption{\label{fig:quality-work}%
  Work needed to get a solution of given quality using star contractions and MST+.
  (Y-axis is fraction of total maximum
  work done and work is measured as number of visited vertices during invocations
  of Dijkstra's algorithm, the work done by MST+ is negligible and hence ignored.)
}
\end{figure*}

Independently, driven by our experiments, we propose two variants of modifications of Zelikovsky's algorithm (Zelikovsky+, Zelikovsky-; see Section~\ref{sec:earlyTerminationAlgs} for their description).
These are based mainly on our insight that takes advantage of reformulating known algorithms in terms of star contractions.
The key idea is to compute stars after each contraction as opposed to precomputing them.
Of course, since we recompute star to contract, the proposed variants are slower (roughly~3 times), however, they both achieve much better performance (i.e., total weight of the solution found); see Figure~\ref{fig:aggregatedMST}.
As it follows from the paragraph above, relaxed star contractions are not helping much since there are not so many large stars.
A large improvement is achieved by iterated MST algorithm (MST+).
This surprisingly cheap, on average approximately 3 times slower than MST, modification of MST yields a significant improvement in approximation.
Consult Figures~\ref{fig:aggregatedMST}, \ref{fig:work} and \ref{fig:Rect} where one can compare MST+ with MST and our modification of Zelikovsky's algortihm.

\paragraph{Our Experiment.}
In what follows we refer to Algorithm~\ref{alg:highLevelOverview}.
We run the algorithm in steps---invocations of the while-loop and every time we call the \Contract{} function we find a solution using all of the aforementioned methods.
We collect the data and, e.g., for MST on public instances from PACE Challenge 2018 we aggregate statistics (see Figure~\ref{fig:aggregatedMST}).
Furthermore, we measure the sizes of contracted stars and performance of \BestStar{}, since this is the most time-consuming step in the algorithm.
It is worth pointing out that we implement some standard heuristics (see Section~\ref{sec:usedHeuristics}) which we use to preprocess the input, that is, all our data is collected on the already preprocessed instances.

Our experiments suggest that the iterated MST (MST+, see Section~\ref{sec:earlyTerminationAlgs}) performs significantly better than classical MST-approximation, even though the time overhead in only multiplicative by a small constant.
This means it is negligible in practice, since MST is computed within seconds on the current inputs.
Compared to the classical MST-approximation, MST+ performs also well combined with star contractions.
We provide all the aforementioned comparisons in Section~\ref{sec:PACEInstances}.

\smallskip
\noindent{\sffamily\textbf{Dataset.}}~~
We evaluate the algorithm and present our results for the set of public instances of PACE Challenge 2018 in \S~\ref{sec:PACEInstances}.
According to the report from PACE Challenge 2018~\cite{PACE-report}, the set of instances in Track C consists of the hardest instances of Steinlib and from real-world telecommunication networks by Ivana Ljubic’s group at the University of Vienna.
It should be noted that similar sources were used for the DIMACS Challenge~\cite{Dimacs}.
By that time, a majority of the selected instances cannot be solved within one hour by the state-of-the-art program and in several cases, the actual optimum was unknown.
For more discussion about the chosen dataset, please consult the report from the PACE Challenge 2018~\cite{PACE-report}.
Furthermore, we evaluate it on rectilinear instances from ORLib~\cite{ORLib} in Appendix~\ref{sec:rect}.

\smallskip
\noindent{\sffamily\textbf{Preliminaries.}}~~
We give a brief recapitulation of graph theory terminology used in this work; for the basic notation, we refer the reader to monographs~\cite{kapitoly, diestel}.
All graphs are undirected without loops and multiple edges.
If we argue about algorithmic complexity of certain routine or the amount of memory needed in order to store some data for a graph $G$, by $n$ we denote the number of vertices of $G$ and by $m$ we denote the number of edges of $G$.
For an edge weighted graph $G = (V,E)$ the \emph{metric closure of $G$}, denoted $\mc(G)$, is the complete graph with the vertex set $V$ with weight of an edge $\{ u,v \}$ equal to the length of the shortest path between $u$ and $v$ in $G$.

For a graph $G = (V, E)$ and an edge $e = \{ u,v \}$ if we \emph{contract} $e$ (denoted as $G/e$), we create a new graph with vertex set $(V \setminus \{ u,v \}) \cup \{ z \}$, where $z$ is a newly introduced vertex.
The edge set of the resulting graph consists of all edges in $E$ not incident to any end-vertex of $e$ together with the newly introduced edges $\{ w,z \}$ for every edge $\{ w,u \}$ as well as for every edge $\{ w,v \}$, where the weight of a newly created edge is the same as the weight of the edge $\{ w,u \}$, $\{ w, v \}$, respectively.
Note that if the above operation is about to create multiple edges we simply keep the one with a lower weight.

For a graph $G = (V, E)$ and a vertex $v$ with exactly two neighbors in $G$ by \emph{suppressing} $v$ we mean changing $G$ into a new graph as follows.
The new vertex set is $V \setminus \{ v \}$ and we delete all edges incident to $v$.
Finally, we edge $\{ x,y \}$ if both $\{ x,v \}, \{ v,y \} \in E$ with weight $w(\{ x,y \}) = w(\{ x,v \}) + w(\{ v,y \})$.

\subsection{More Details on Past Implementation Challenges}
Since \textsc{Steiner Tree} has many applications, it received attention among practitioners and in operations research.
One particular example can be e.g.\ the specialized module SCIP-Jack~\cite{GamrathKMRS17} in the SCIP tool for solving (mixed) integer linear programs.
In the 11\textsuperscript{th} DIMACS Implementation Challenge~\cite{Dimacs}
various variants of \textsc{Steiner Tree} formed the central topic of the challenge.
Among others, e.g.\ the basic version \textsc{Steiner Tree}, geometrical versions (e.g., rectilinear instances), and prize-collecting variants were tackled.
One can read in the description of the DIMACS Implementation Challenge:
\begin{quote}\emph{%
DIMACS Implementation Challenges address questions of determining realistic algorithm performance where worst-case analysis is overly pessimistic and probabilistic models are too unrealistic: experimentation can provide guides to realistic algorithm performance where analysis fails.
}\end{quote}
Last but not least, one track of the PACE Challenge 2018~\cite{PACE-report,PACE-web}
was completely devoted to \textsc{Steiner Tree} with three specialized branches.
In PACE Challenge2018 there was an approximation branch and two exact branches---one with the additional promise of a small number of terminals and in the other, a tree-decomposition of the input graph of small tree-width was given.

In practice, the implementation of star contraction algorithm~\cite{HTKME} accompanied with few more tricks is quite competitive with other established solutions.
Table~\ref{tab:pace} compares the performance of their algorithm with other algorithms participating in the PACE Challenge 2018.
It is important to note that their algorithm uses a randomized local search and improved MST (see Section~\ref{sec:earlyTerminationAlgs} for a description of MST+) in addition to star contractions.
For more details of their implementation, consult the repository with the detailed description~\cite{HTKME}.
In the challenge, a time-limit to output a solution was set to 30 minutes---we call this a run.
Afterward, each run received points according to the fraction of the value of the returned solution to the best solution known.
Those were aggregated over all instances.

\begin{table}
  \centering
  \begin{tabular}[bt]{lc}
    \hline
    \multicolumn{1}{c}{Algorithm} & Score \\ \hline
    Evolution Algorithm & 99.91\\
    MIP solver + Heuristics (SCIP-Jack~\cite{SCIP}) & 99.89\\
    Iterated Local Search & 99.78\\
    \textbf{Star Contractions + Local Search~\cite{HTKME}} & \textbf{99.70} \\
   Zelikovsky~\cite{Zelikovsky93} & 98.93\\
   Simulated Annealing & 98.27\\
   Random Generation + Local Search & 97.54\\
   Shortest Path Heuristic$^*$
   + Local search &97.15 \\
   Mehlhorn 2-approximation~\cite{Me88} + & \\  Watel and Weisser $k$-Approximation for & \\ the directed Steiner Tree Problem~\cite{WaWe16}  & 96.92 \\
   Shortest Path Heuristic$^*$ & 94.57 \\
   Primal-Dual 2-approximation + Local Search  & 94.37 \\
   Contract Random 2-terminal Shortest Path & \\ + MST &  82.61 \\
   Ant Colony Optimization & 80.73 \\
   \hline
  \end{tabular}
\medskip
  \caption{%
    An aggregated comparison of the overall performance of the algorithms participating in PACE Challenge 2018 Track C.
    We exchange the names of the teams with a summarizing name of the main method they used.
    For more implementation details of the algorithms see the report~\cite{PACE-report}, which also contains links to individual implementations and comprehensive descriptions.
\newline
    $^*$\textit{Shortest Path Heuristic}: Pick one terminal as a root and repeat the following: Find the closest terminal to the root and contract the shortest path from this terminal to the root.
  }\label{tab:pace}
\end{table}

\section{Implementation Details and Heuristics}\label{sec:algorithmDetails}
In this section we describe our implementation and improvements of finding the Best Star as proposed in~\cite{DvorakFKMTV18} (\S~\ref{sec:findingBestStar}), the approximate algorithms used to finish the solution after the application of the Best Star Algorithm (\S~\ref{sec:earlyTerminationAlgs}), and also the heuristics we used to preprocess the instances (\S~\ref{sec:usedHeuristics}).
Last but not least, we discuss a few simple yet in practice well-performing modification of Zelikovsky's algorithm (\S~\ref{sec:earlyTerminationAlgs}).

\subsection{Used Heuristics}\label{sec:usedHeuristics}
All heuristics we used are deterministic and ensure that the optimal value of instance before and after applying them is the same (such heuristics are called 1-safe in the so-called lossy kernelization framework~\cite{lokshtanov2017lossy}).
Namely, we used the following well-known ones:
\begin{enumerate}
  \item\label{it:cheapHeuBegin} We contract all edges $e$ with $w(e) = 0$ at the very beginning\footnote{It is worth noting that in this case the contracted edge is not guaranteed to be a part of the solution (there may be an isolated component containing only some of these edges). Thus, in the end, we have to remove all components of the computed solution that do not contain any terminal.}.
	Thus we assume in the rest that all weights are positive.
  \item We remove all Steiner vertices of degree 1 and suppress Steiner vertices of degree 2.
  \item We contract edges incident to terminals of degree 1.
  \item\label{it:cheapHeuEnd} Contract an edge $e = \{ s,t \}$ between two terminals if $w(e)$ is minimal among edges incident to $s$ or $t$.
  \item {\em Shortest Path Test (SPT):} Delete an edge $e = \{ u,v \}$ if $w(e)$ exceeds the length of the shortest path between $u$ and $v$.
  \item {\em Terminal Distance Test (TDT)} \sv{We postpone the description to the appendix, Section~\ref{app:tdt}.}\lv{see below.}
\end{enumerate}

\SetKwFunction{BuyZero}{contract\_zero\_edges}
\SetKwFunction{DegOne}{delete\_degree\_one\_Steiner}
\SetKwFunction{TerminalOne}{contract\_the\_only\_edge\_incident\_to\_term}
\SetKwFunction{TerminalCheapest}{contract\_the\_cheapest\_edge\_between\_two\_terminals}
\SetKwFunction{CheapHeu}{quick\_heuristics}
\SetKwFunction{SPTShort}{SPT\_two\_edges}
\SetKwFunction{SPTLong}{SPT}
\SetKwFunction{TDT}{TDT}
\SetKwFunction{Heu}{preprocessing}
\toappendix{
\begin{algorithm2e}[h!]

\Pn{\CheapHeu{$G$}}{
  $run \assign \texttt{true}$ \;
  \While{run}{
    $run \assign \texttt{false}$\;
    $run \assign$ \BuyZero{$G$}\;
    $run \assign run \lor \DegOne{$G$}$\;
    $run \assign run \lor \TerminalOne{$G$}$\;
    $run \assign run \lor \TerminalCheapest{$G$}$\;
  }
}

\Pn{\Heu{$G$}}{
  \CheapHeu{$G$} \;
  \SPTLong{$G$} \;
  \CheapHeu{$G$} \;
  \TDT{$G$} \;
  \CheapHeu{$G$} \;
  \SPTLong{$G$} \;
  \CheapHeu{$G$} \;
}

\medskip
\caption{\label{alg:heuristicRun}%
Pseudocode of the used preproscessing.
}
\end{algorithm2e}
}

\toappendix{%
Heuristics (\ref{it:cheapHeuBegin}) up to (\ref{it:cheapHeuEnd}) are implemented using straightforward iteration over all edges or vertices of the graph, and if any of them succeeds, we again rerun all of these.
This iteration blows up theoretical time complexity by a factor of $n$ but in practice, only very few iterations yield an irreducible instance.
In Algorithm~\ref{alg:heuristicRun} we encapsulate these into the \CheapHeu{} function.
For performance reasons we split the SPT heuristic into two parts: first, it checks only paths consisting of two edges which is quite efficient ($O(n^2)$ because we store edges incident to every vertex in a sorted order) and then we check paths of all lengths which requires to run Dijkstra's algorithm~\cite{Dijkstra59} from every vertex and is therefore much slower.

It is worth noting that both SPT and TDT can benefit from rerunning but due to their time complexity and usually lower number of improvements, we do not use these heuristics exhaustively.
Instead, we run SPT, TDT, and SPT once more and execute \CheapHeu{} at the beginning, between them,
and at the end.
See the \Heu{} function in Algorithm~\ref{alg:heuristicRun}.
}

\toappendix{
\lv{\noindent{\sffamily\textbf{Terminal Distance Test.}}~~}
The TDT heuristic was introduced in~\cite{heu:TDT}.
The basic idea is the following: Let $(W, W')$ be a partition of vertices
such that both $W$ and $W'$ contain some terminal and each of them is connected,
let be $e$ and $f$ the shortest and the second shortest edge of the cut induced by
$(W, W')$. If there is a path connecting some terminal $t \in W \cap T$ and
$t' \in W' \cap T$ which uses $e$ and is no longer that $f$ then there exists an
optimal solution which uses edge $e$.

Note that while it is easy to verify
the correctness of this heuristic, it does not give us directly an effective algorithm.
Koch and Martin~\cite{KochM98} point out that it is possible to implement TDT in time $O(|V|^3)$.
Furthermore, they claim that the time complexity can be further improved to $O(|V|^2)$ as is described in the thesis~\cite{DuinPhD-TDT}, however, we were unable to access the thesis.
Consequently, we describe our implementation in Algorithm~\ref{alg:tdt} \sv{ (See Appendix~\ref{app:algorithmDetails})}, as we hope it might be interesting for future reference. Note that the algorithm presented here
intentionally omits handling cases where multiple edges have the same weight.

Our implementation is based on the data structure for dynamic edge 2-connectivity
of Westbrook and Tarjan~\cite{Westbrook1992}. This structure has amortized time
complexity $O(\alpha(m))$ per operation and supports edge addition
and check whether two vertices belong to the same (2-connected) component.
The time complexity of Algorithm~\ref{alg:tdt} is $O(nm \log n)$ when
Dijkstra's algorithm is implemented with $d$-regular heap.
}

\toappendix{
\begin{algorithm2e}[h!]
\SetKwFunction{TDT}{TDT}
\SetKwFunction{sort}{sort}
\SetKwFunction{addedge}{add\_edge}
\SetKwFunction{testedge}{test\_edge}

\Pn{\testedge{$G$, $e$, $f$, $X$}}{
  $u, v \assign e$ \;
  Remove $e$ from $G$ \;

  $t_1 \assign \textrm{terminal closest to } u$ \tcp*{Dijkstra's algorithm}
  $t_2 \assign \textrm{terminal closest to } v$ \;

  \If{$d(t_1, u) + w(e) + d(v, t_2) \leq w(f)$}{
    $X \assign X \cup e$ \;
  }

  Add $e$ back to $G$ \;
}

\Pn{\TDT{$G$}}{
  $E' \assign \sort{$E(G)$, $w(a) < w(b)$}$\;
  $C \assign \textrm{structure for 2-connectivity}$\;
  $X \assign \emptyset$ \;

  \ForEach{$e \in E'$}{
    $B \assign \addedge{$C$, $e$}$
    \tcp*{returns edges which were bridges before addition of $e$ but no longer are}
    \ForEach{$b \in B$}{
      \testedge{$G$, $b$, $e$, $X$} \;
    }
  }

  Buy edges in $X$ \;
}

\medskip
\caption{\label{alg:tdt}%
Implementation of Terminal Distance Test}
\end{algorithm2e}
}

\subsection{Finding the Best Star}\label{sec:findingBestStar}
As already mentioned in \S~\ref{sec:intro}, Algorithm~\ref{alg:highLevelOverview} repeatedly finds and contracts a so called best star (according to a certain ratio) in $G$.
Here, we first give the definition used by Dvo{\v{r}}{\'{a}}k et al.~\cite{DvorakFKMTV18}.
Later, we discuss a further additional practical extension of the former definition and describe our implementation in detail.

Let $G$ be a graph and $R$ the set of terminals in $G$.
For a vertex $c$ and a set $R' \subseteq R$ we define a \emph{star} (centered at $c$ and a terminals set~$R'$) which we denote $\star(c, R')$.
We define a \emph{ratio of star} $\star(c, R')$ as
\[
\starRatio\left( \star(c, R') \right) = \frac{\sum_{t \in R'} \dist(c, t)}{|R'| - 1},
\]
where $|R'| \ge 2$ and $\dist(c,t)$ is the weight of the edge $\{ c,t \}$ in $\mc(G)$.
The best ratio achievable for a star centered at $c$ is $\starRatio(c) = \min_{R' \subseteq R,\, |R'| \ge 2} \starRatio(\star(c, R')) $ and a \emph{best star centered at $c$} is any minimizer of the defining expression.
The \emph{best star in the graph $G$} is any star $\star(c,R')$ minimizing $\starRatio(G) = \min_{c \in V} \starRatio(c)$.

Clearly, the smaller the ratio the better.
On the other hand, in our experiments, there were many ties and thus we further extend this definition in the case there are more minimizers of the best ratio in $G$.
We introduce a second measurement taking into account the number of terminals contained in a star.
Intuitively, the more terminals it contains the better.
Thus the \emph{best star in the graph $G$} is any star $\star(c,R')$ with ratio $\min_{c \in V} \starRatio(c)$ which maximizes $|R'|$.

It is worth noting that if a best star with center $c$ contains $k$ terminals, then it contains of the $k$ terminals closest to $c$ in $\mc(G)$~\cite[Lemma~6]{DvorakFKMTV18}.
Even though the definition of the best star uses $\mc(G)$, we cannot afford to compute and store it due to its size (quadratic in $n$).
Instead, we utilize Dijkstra's algorithm~\cite{Dijkstra59} to compute the best star centered in a given vertex.
In total we obtain running time $O(mn + n^2 \log n)$ per one round, that is, for one execution of the main loop in Algorithm~\ref{alg:highLevelOverview}.
Furthermore, we use several heuristics to improve the running time significantly in practice.
These heuristics employ memorization and early termination among others; refer to Algorithm~\ref{alg:findBestStar}\sv{ in Appendix~\ref{app:algorithmDetails}}.
Let $\starRatioCur(G)$ denote the best already computed ratio (i.e., the best ratio among all already computed stars).
By slightly overloading the notation when searching for the best star centered at $c$ we let $\starRatioCur(c)$ denote the ratio of the best so far computed star centered at $c$ (in the current invocation of the search).
Namely:
\begin{enumerate}
  \item
    We stop the execution of Dijkstra's algorithm when the current distance from the source (center of the star) is strictly greater than $\starRatioCur(c)$.
  \item
    For every vertex $c \in V$ we store a best star centered at $c$ in between the rounds.
  \item
    We compute the best star at $c$ only if the stored one could have been
    affected by a star-contraction performed in the previous round.
    We can do this since a single star-contraction affects only a small (local) part of the graph.
\end{enumerate}

\smallskip
\noindent{\sffamily\textbf{Overestimating Star Weight.}}~~
It is worth noting that best star as described by Dvo{\v{r}}{\'{a}}k et al.\ is a star in metric closure containing some number of terminals closest to its center (and minimizes the star ratio).
Note that this clearly may overestimate the weight of such a star as well as its ratio (see Figure~\ref{fig:overestimatingWeight}\sv{ in the appendix}).%
\toappendix{%
\begin{figure}[bt]
  \begin{center}
    \includegraphics{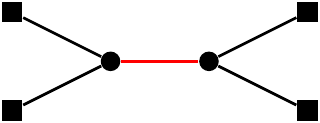}
  \end{center}
  \caption{\label{fig:overestimatingWeight}%
  Simple graph containing four terminals (represented by boxes) and two Steiner vertices (discs).
  Suppose all the edges have unit weight.
  When computing the weight of a star centered at any of the two Steiner vertices using the metric closure, we count the weight of the red edge twice.
  As a consequence, one gets two possible stars with ratio 2---one containing only the two terminals connected directly to the assumed Steiner vertex and the other containing all four terminals.
  However, the second described star should better have a ratio of $5/3$ (which is better than 2).
  }
\end{figure}%
}
Observe that if the best found star contains at most three terminals, then its weight is always estimated precisely.
While this estimate is sufficient for the purpose of theoretical analysis, it may affect the overall behavior of the algorithm.
We would like to point out that in our implementation of a star-contraction if the best star in $\star(c, R')$, we contract edges of an MST containing $R'$ instead of contracting that star itself.
Clearly, this modification can only decrease the cost in one single round.
On the other hand, it is not clear how this ``greedy'' improvement affects the overall performance of the whole process.
\lv{In the next paragraph, we propose a way to overcome the overcounting issue.}

\smallskip
\noindent{\sffamily\textbf{Improved stars.}}~~
Using Dijkstra's algorithm we are recursively searching a new terminal within threshold distance and building the best star for each vertex.
Whenever we encounter a terminal such that it forms a better star with currently best star originating in the given vertex we add it to the constructed star and start over setting the whole star as the origin for Dijkstra's algorithm.

\toappendix{
\begin{algorithm2e}[h!]
\SetKwFunction{BestStar}{find\_best\_star}
\SetKwFunction{BestStarVertex}{find\_best\_star\_v}
\SetKwFunction{Dijkstra}{dijkstra\_next}
\SetKwFunction{Lowerbound}{lbound}
\SetKwFunction{RatioInvalid}{invalid}

\Pn{\BestStarVertex{$G$,$R$,$v$,$r$}}{
  $weight \assign 0$     \tcp*{sum of distances}
  $nter \assign 0$       \tcp*{current number of terminals}
  \While{$(w,d) \assign \Dijkstra{G,v}$}{
    \lIf{ $nter \ge 2 \land weight / (nter-1) < d$ }{
      \Return{ $weight / (nter-1)$ }
    }
    \lIf{$d > 2r$}{
      \Return{ \Lowerbound{$d$} }
    }
    \If{$w \in R$}{
      $weight \assign weight + d$ \;
      $nter \assign nter + 1$ \;
    }
  }
}

\Pn{\BestStar{$G$,$R$}}{
$\starRatioCur(G) \assign \infty$ \;
  \lForEach{$v\in V$}{
  $ratio[v] = \infty$ 
  }
  \ForEach{$v\in V$}{
    \If{$\RatioInvalid{v} \lor { (ratio[v] < \starRatioCur(G) \land \Lowerbound{$ratio[v]$} ) } $}{
      $ratio[v] \assign \BestStarVertex{G,R,v,$\starRatioCur(G)$}$ \;
    }
    \If{$\starRatioCur(G) > ratio[v]$}{
      $\starRatioCur(G) \assign ratio[v]$ \;
      $star\_center \assign v$ \;
    }
  }
  \Return{ $(star\_center, \starRatioCur(G))$ } \;
}

\medskip
\caption{\label{alg:findBestStar}%
A pseudocode for The Best Star function.
}
\end{algorithm2e}
}

\subsection{Finishing the partial solution}\label{sec:earlyTerminationAlgs}
The original algorithm of Dvořák et al.~\cite{DvorakFKMTV18} performs star
contractions until the number of terminals decreases under a threshold depending
on the desired approximation ratio $\varepsilon$, and then uses an exact algorithm.
While this is a very natural theoretical approach, it is not suitable for practical use for the following reasons:
\begin{enumerate}
\item Both discussed exact algorithms
are based on the dynamic programming which makes them quite slow in practice
(virtually intractable for instances with more than 20 terminals).
\item The threshold
depends on the number of Steiner vertices in (some) optimal solution, and we
in general have no good upper bound for it.
\end{enumerate}

Instead, for the purposes of the evaluation, we choose the following algorithms which we run after every contraction:

\toappendix{
  \begin{algorithm2e}[t]
  \SetKwFunction{Zelikovsky}{zelikovsky}
  \SetKwFunction{Voronoi}{voronoi\_partition}
  \SetKwFunction{Dijkstra}{dijkstra}
  \SetKwFunction{MST}{MST}

  \Pn{\Zelikovsky{$G$, $R$}}{
    $G' \assign G$ \;

    $d[\cdot] \assign \infty$ \;
    $c[\cdot] \assign \texttt{None}$ \;
    $vor \assign \Voronoi{$G$, $R$}$ \;
    \ForEach{$v \in V$}{
      $d_v \assign \Dijkstra{$G$, $v$}$ \;
      \ForEach{$u_1, u_2 \in {R \setminus \{v\} \choose 2}$}{
        $w \assign d_v[vor[v]] + d_v[u_1] + d_v[u_2]$ \;
        \If{$w < d[\{ vor[v], u_1, u_2 \}]$}{
          $d[\{ vor[v], u_1, u_2 \}] \assign w$ \;
          $c[\{ vor[v], u_1, u_2 \}] \assign v$ \;
        }
      }
    }
    \BlankLine

    $S \assign \emptyset$ \;
    \While{{\tt True}}{
      $t_b \assign \texttt{None}$ \;
      $win \assign 0$ \;

      \ForEach{${t} \in {R \choose 3}$}{
        \lIf{$|{t}'| \neq 3$}{\Continue }

        $w \assign \MST{$G'$} - \MST{$G' / {t}'$} + d[{t}]$
          \tcp*{$x'$ is the set of vertices of $G'$ corresponding to vertices $x$ of $G$}

        \If{$win < w$}{
          $win \assign w$ \;
          ${t}_b \assign {t}$ \;
        }
      }

      \lIf{$win = 0$}{\Return{\MST{$G$, $R \cup S$}}}

      $G' \assign G' / {t}_b'$ \;
      $S \assign S \cup \{ c[{t}_b] \}$ \;
    }
  }

\medskip
  \caption{\label{alg:zel}%
    Implementation of the original Zelikovsky's algorithm according to~\cite{Zelikovsky93}.
}
\end{algorithm2e}
}
\begin{itemize}
  \item {\bf MST:} The usual well-known minimum spanning tree approximation---we take a subgraph of the metric closure induced by terminals and find its minimum spanning tree.
    Note that the implementation does not compute whole metric closure but computes Voronoi regions of the terminals instead.
    Then it constructs auxiliary graph using terminals of the original graph as vertices and adding an edge for every edge $\{u,v\}$ crossing between Voronoi regions with length $d(u) + w(u, v) + d(v)$ where $w(\cdot,\cdot)$ is length of an edge and $d(\cdot)$ is distance to closest terminal~\cite{EuclidianVoronoi}.
  \item {\bf MST+:} We calculate MST and then improve its solution by taking its terminals and branching vertices (Steiner vertices with the degree at least 3 in the solution), marking
    them all as terminals and running MST on this modified instance. It is easy
    to see that solution of MST on this modified instance is
    never worse than the solution we began with because
    the original solution is a spanning tree of the modified instance.
    We repeat this while the solution is improving.
  \item {\bf Zelikovsky:} The Zelikovsky's algorithm~\cite{Zelikovsky93}, which
    augments the MST solution using stars with 3 terminals, was the first
    algorithm with a better approximation ratio than 2.
    The algorithm proceeds in rounds. Each round it looks at all 3-stars, selects the star $s$ which maximizes the so-called ``$win$'' which is $mst(G) - mst(G / s) - d(s)$
    (where $mst(G)$ is the weight of MST solution of instance $G$, $G / s$ denotes
    $G$ after contraction of terminals in $s$ and $d(s)$ is the weight of star $s$),
    and if the $win$ of the best star is strictly positive, it contracts it and
    starts another round, otherwise, it stops, and returns MST on all terminals
    and selected star centers. (Computing the MST at the end is needed to ensure
    that the solution is really a tree.)
    The original version of the algorithm computes the best center for every triple
    of terminal and weight of such a star at the very beginning and
    runs in $O(n(m + n \log n + t^2) + t^4)$ time and requires $O(m + t^3)$ extra space.
    You can see the pseudocode in Algorithm~\ref{alg:zel}\sv{ (see Appendix~\ref{app:algorithmDetails})}.
  \item {\bf Zelikovsky$-$:}
    This modification recomputes distances of triples in each round instead of precomputing them in advance (ln.\ 6--12 in Algorithm~\ref{alg:zel}).
    Unlike the usual Zelikovsky's algorithm where the triplets are only upper bounds, here we have optimal values in each round.
   This algorithm slower ($O(nt(m + n \log n + t^2))$)
    but the memory requirement is $O(m)$ smaller.
  \item {\bf Zelikovsky+:}
    This modification differs from Zelikovsky$-$ only by the application of MST+ where MST instead of MST at the very end.
\end{itemize}

\begin{example}[Zelikovsky-]\label{ex:zelikovsky}
  The Zelikovsky's algorithm also fits into the star-meta-algorithm framework
  described in the introduction:
  The parameters are $\tau = 2$ and $k = 3$, the \Eval{} function returns $-win$
  (or $\infty$ for nonpositive $win$), the \Contract{} just contracts the star
  and adds center of the star to $S$,
  and the \Finish{} computes MST of $R \cup S$. Note that Zelikovsky$-$ variant is
  more natural as it finds the best star each round whereas the original Zelikovky's algorithm precomputes values of all stars with 3 terminals.
\end{example}

\section{Outcomes of our Experiments}\label{sec:PACEInstances}
We have implemented the aforementioned approximation algorithms and the star contraction algorithm framework.
We have tested our implementation and provided statistics of the algorithms' performance.
We conclude that the framework constitutes an interesting view on the \textsc{Steiner Tree} problem, however, more experiments are probably still to be done.
This is mainly because too much work is done in the algorithm phase when computing the new best star to be contracted.
We identify the main disadvantage of the algorithm of Dvořák et al.~\cite{DvorakFKMTV18} as the need for maintaining the metric closure of the input graph.
We thus propose a way to avoid this step as well as a heuristic to avoid too much recomputation of the stars found around each terminal.
We stress here that our methods are mainly based on ``star/contraction locality'', that is, if the contracted star could not affect some star, the later one shall still be valid.
Furthermore, in our implementation, it is possible to use caching (which significantly reduced the overall running time), however, this is probably not sufficient for the successful use of the proposed algorithm in practice.

In this section, we perform our main tests that are carried out on the instances\footnote{\url{http://www.lamsade.dauphine.fr/~sikora/pace18/heuristic.zip}}
from the PACE Challenge 2018, Track C.\footnote{It is worth noting that all charts for all instances are available in the repository containing our implementation code
\url{https://github.com/JohnNobody-3af744f30980b7458372/star-contractions}.}\footnote{Due to time constraints all the tests were perfomed only on 123 out of 200 instances from the PACE Challenge 2018.
The problem was running Zelikovsky's algorithm after each best star contraction which took too long on large instances.
We provide figures where tests (excluding Zelikovsky's algorithm) were performed on almost all instances (excluding instances number 193, 196, 197, and 198 only) in Appendix~\ref{app:allInstances}.
Those four instances were still too large for testing.
}
Aggregated data from the measurements are provided in Figure~\ref{fig:aggregatedMST} and the corresponding data in Appendix\lv{~\ref{sec:data}} Table~\ref{tab:aggregatedMST}.
There, star contractions are executed in rounds and in every round, all heuristics (from Section~\ref{sec:earlyTerminationAlgs}) are executed so that the overall performance can be compared.
Experiments are done separately for basic stars and only some of them are repeated for improved stars.
An important thing to note is that the best solution for each input was derived out of the best of outcomes from all the performed experiments improved by an additional local search heuristic.
Therefore, it represents more or less the current state of the art best solution.
Also, the worst solution was obtained using heuristics described in Section~\ref{sec:usedHeuristics}.

The basic outcome of our experiments is the chart for MST.
It shows that the best star algorithm improves significantly the performance of an MST approximation.
Here, the star contractions help to improve the quality of some solutions for more than 57\%.
Despite this was expectable it is good that it is confirmed experimentally.

Our experiments suggest that MST+ should be used whenever MST is usually used.
It outperforms MST in every measurement but still, it is quite simple to code and very fast.
Moreover, star contraction combines very well with MST+ which improves not only the overall performance but more importantly a good solution is obtained much sooner.

\begin{figure*}
  \centering
  \includegraphics[width=0.49\textwidth]{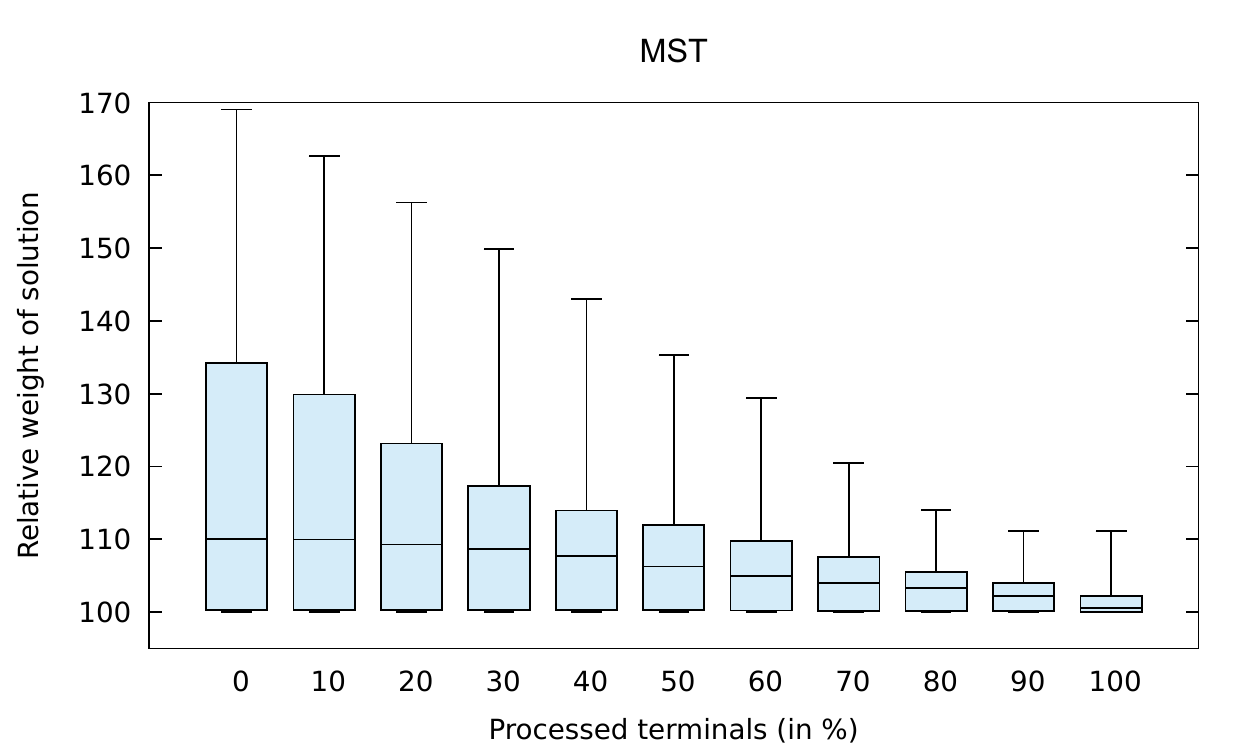}
  \includegraphics[width=0.49\textwidth]{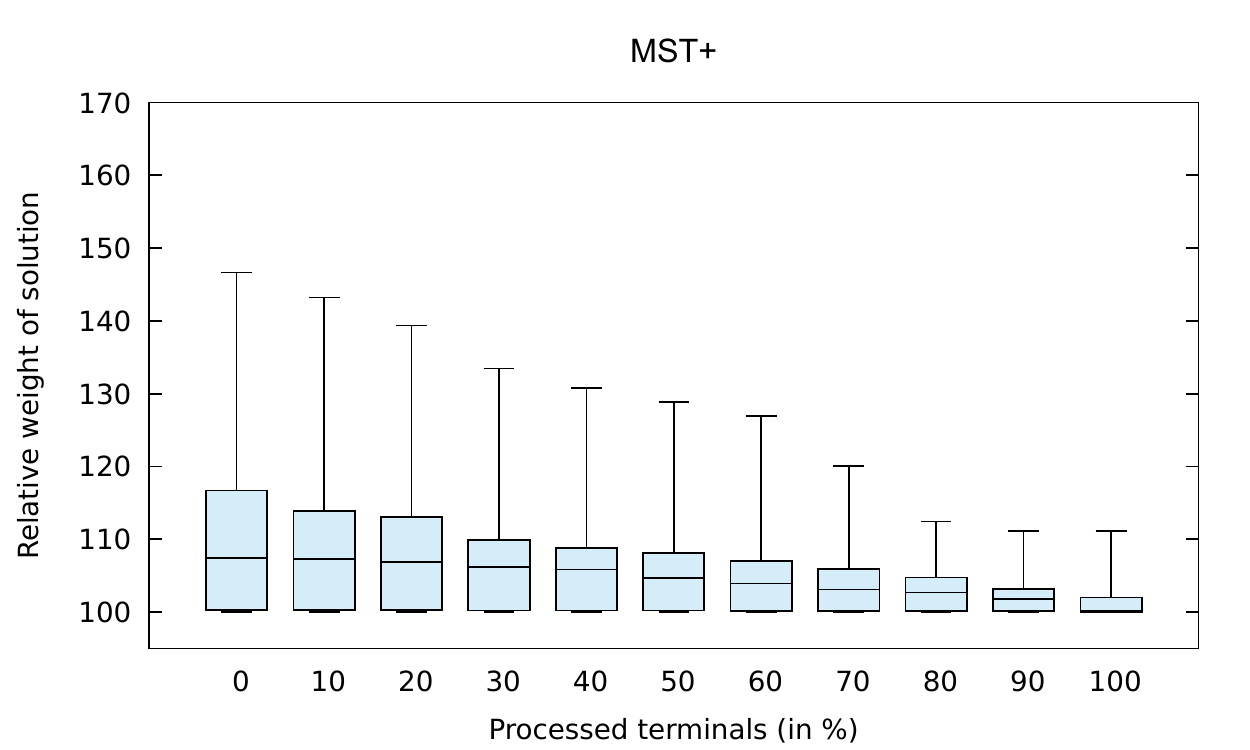}
  \includegraphics[width=0.49\textwidth]{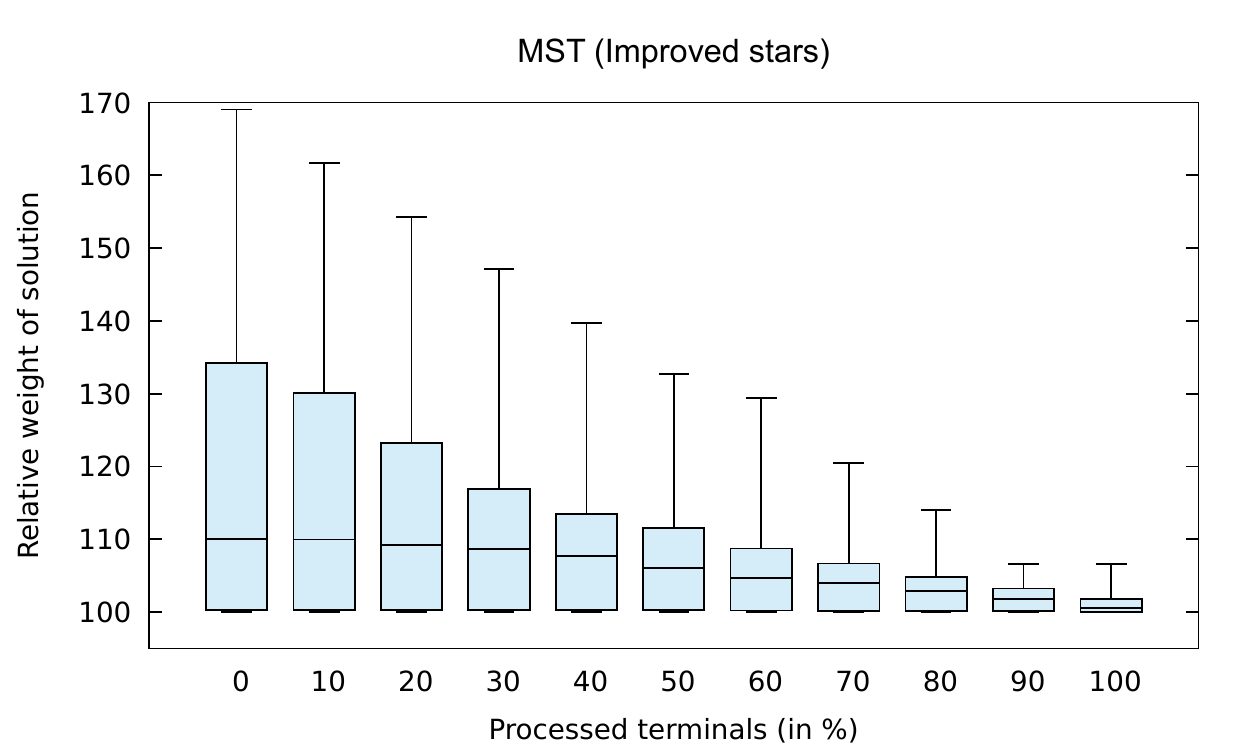}
  \includegraphics[width=0.49\textwidth]{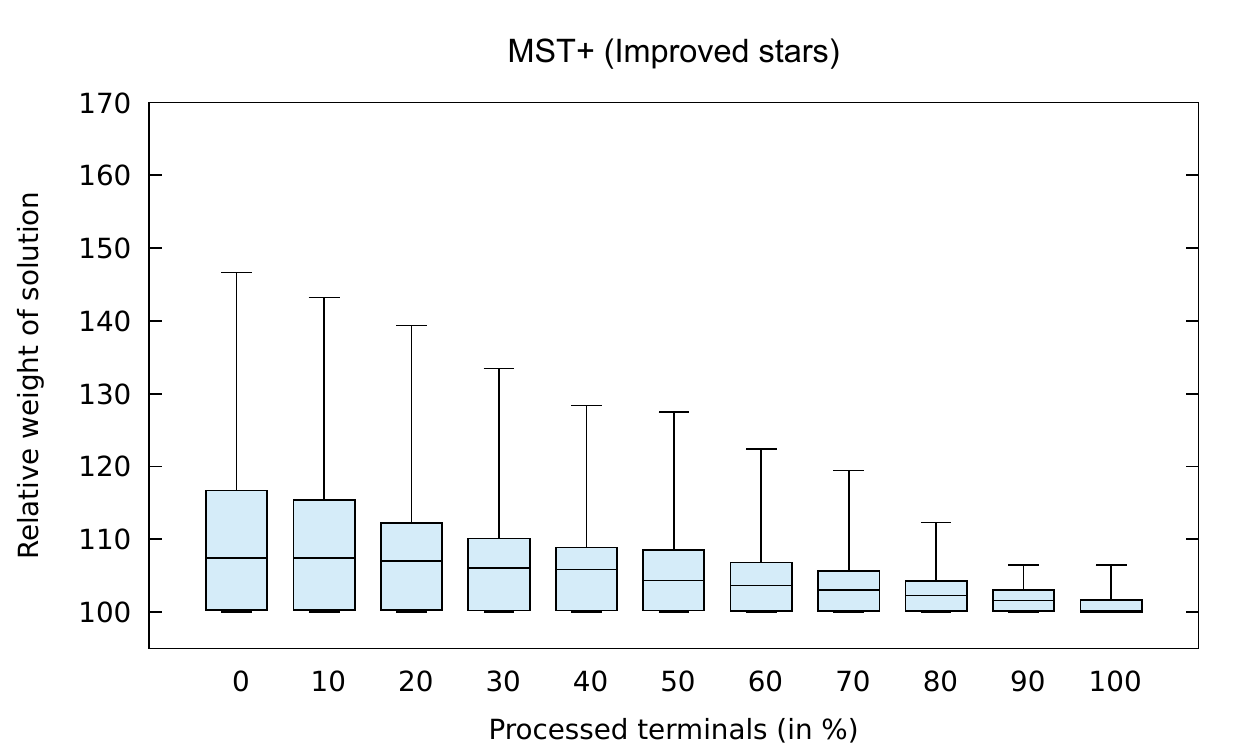}
  \includegraphics[width=0.49\textwidth]{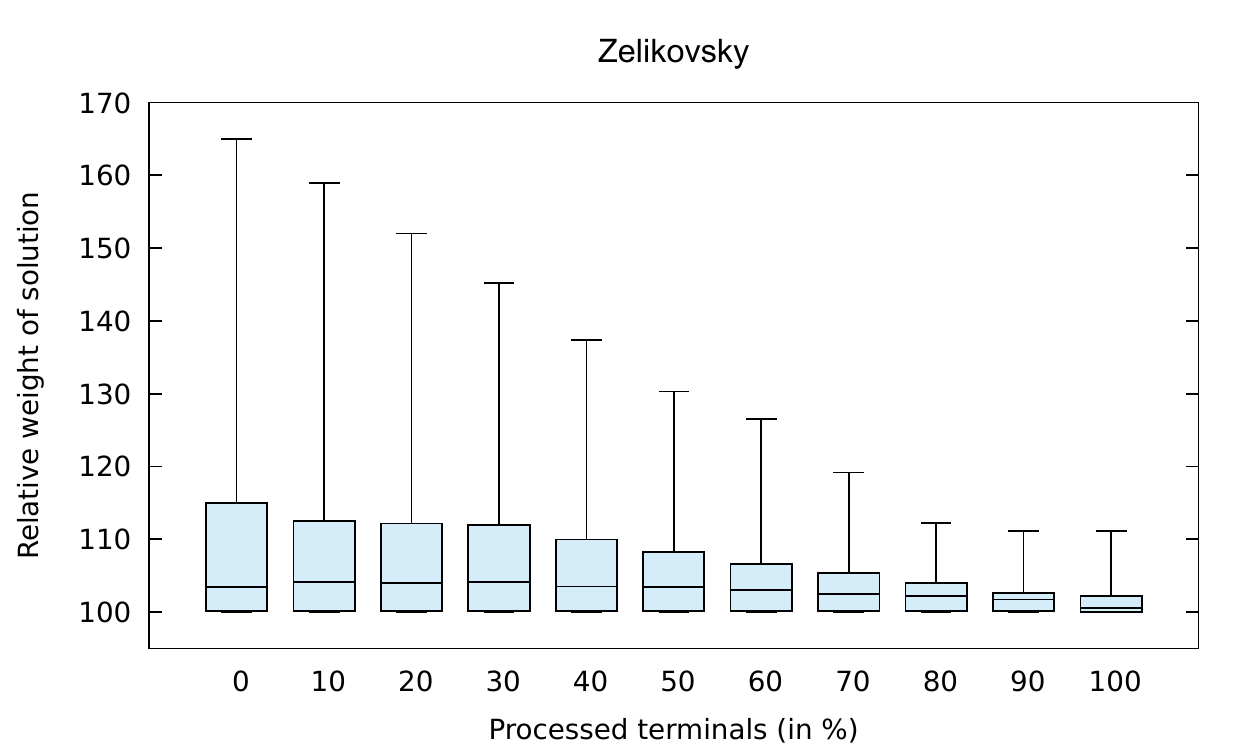}
  \includegraphics[width=0.49\textwidth]{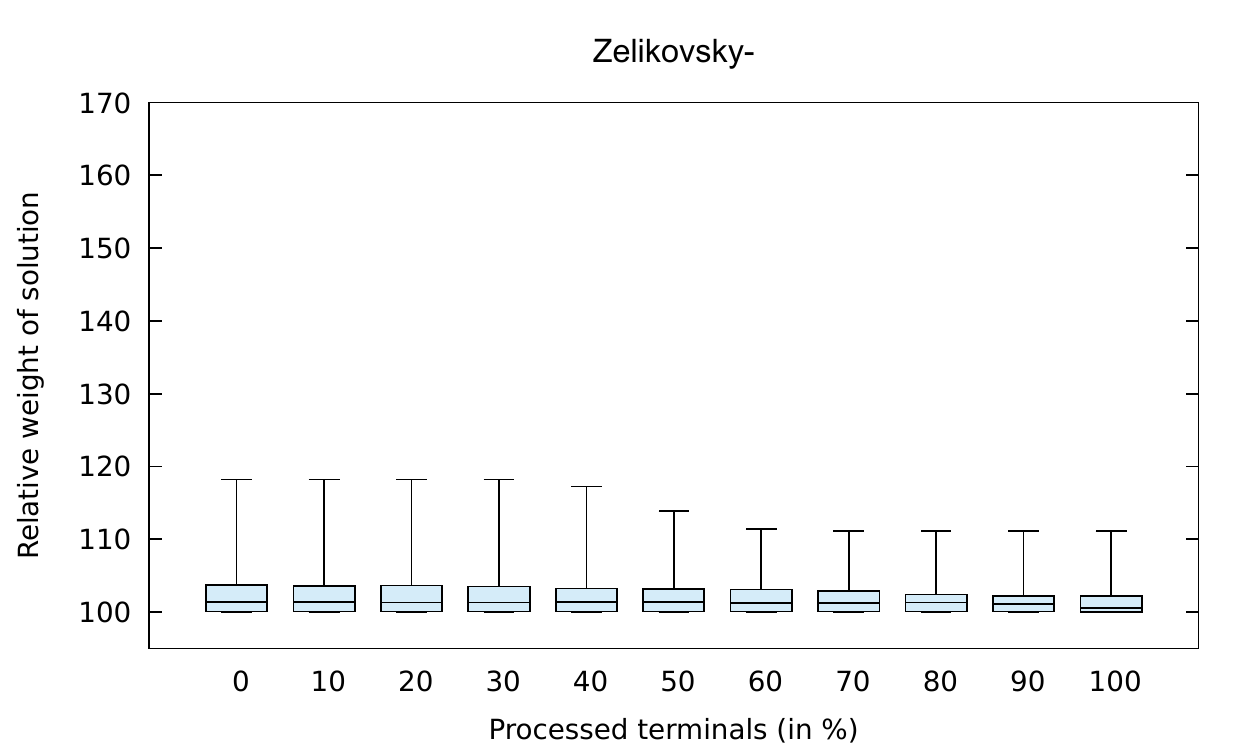}
  \includegraphics[width=0.49\textwidth]{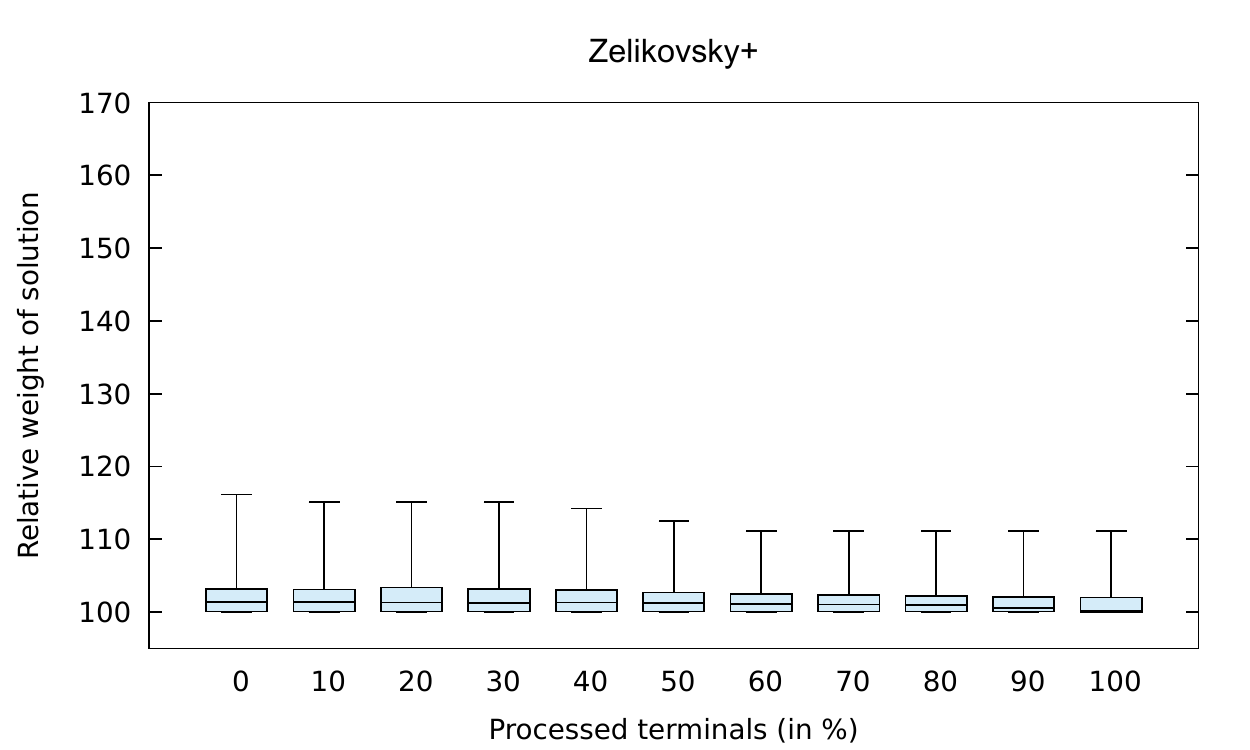}
  \includegraphics[width=0.49\textwidth]{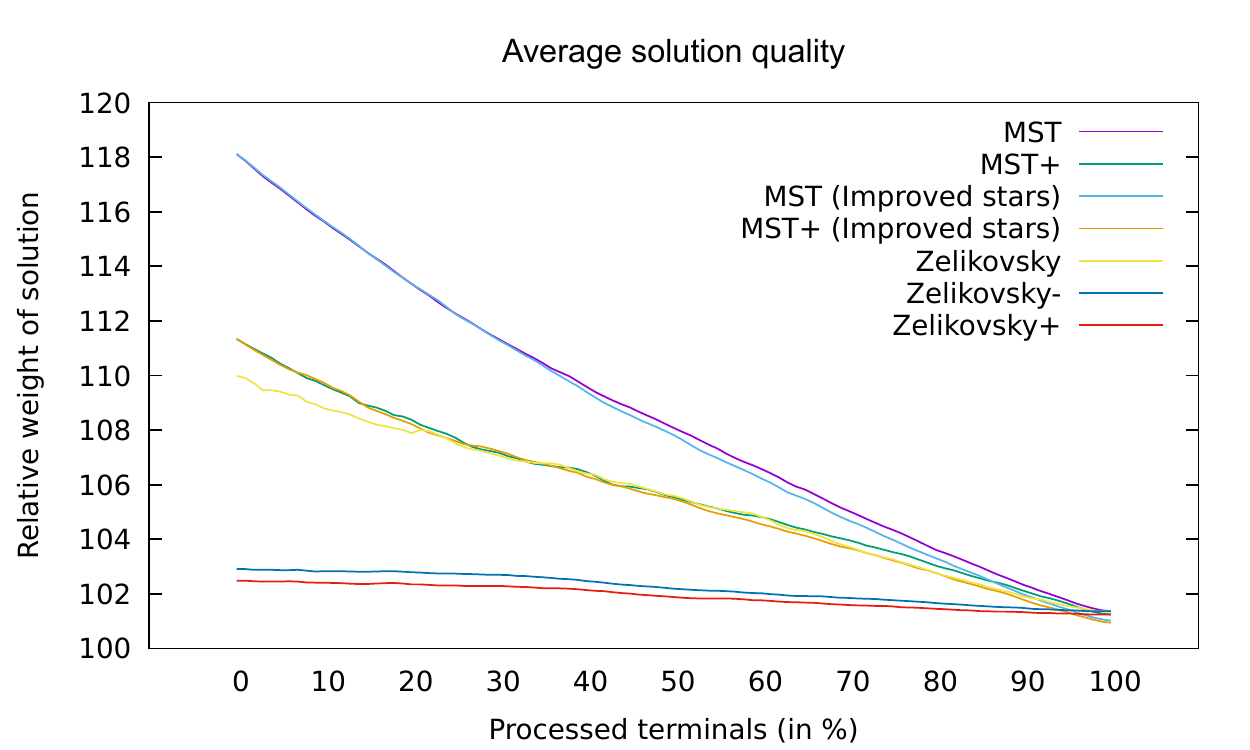}
\medskip
  \caption{\label{fig:aggregatedMST}%
    This chart shows the performance comparison of star contractions and MST heuristics
    on PACE Challenge 2018 instances (for raw data see Table~\ref{tab:aggregatedMST}).
The x-axis represents the number of star contractions in percent before MST was computed.
    The zero value is MST heuristics after preprocessing only.
    Hundred denotes a result obtained by star contractions till one vertex remains in the graph.
 The y-axis represents the quality of the solution again in percentage where the hundred is the best solution we obtained during our experiments (not only in this comparison).
 It is important to point out that the best solution we are comparing to was derived using a local search algorithm, so optima are represented by more or less the current state of the art results.
    The top line is the maximum in our dataset, the colored box represents data points from the first to the third quartile with the line in the middle denoting the median, and the line at the bottom is the minimum.
    The last plot shows arithmetic averages of the same data combined into a single plot for easier comparison.
    (For raw data see Tables~\ref{tab:aggregatedMST} and~\ref{tab:aggregatedMST2}.)
}
\end{figure*}

\begin{figure*}[h!]
  \centering
  \includegraphics[width=0.48\textwidth]{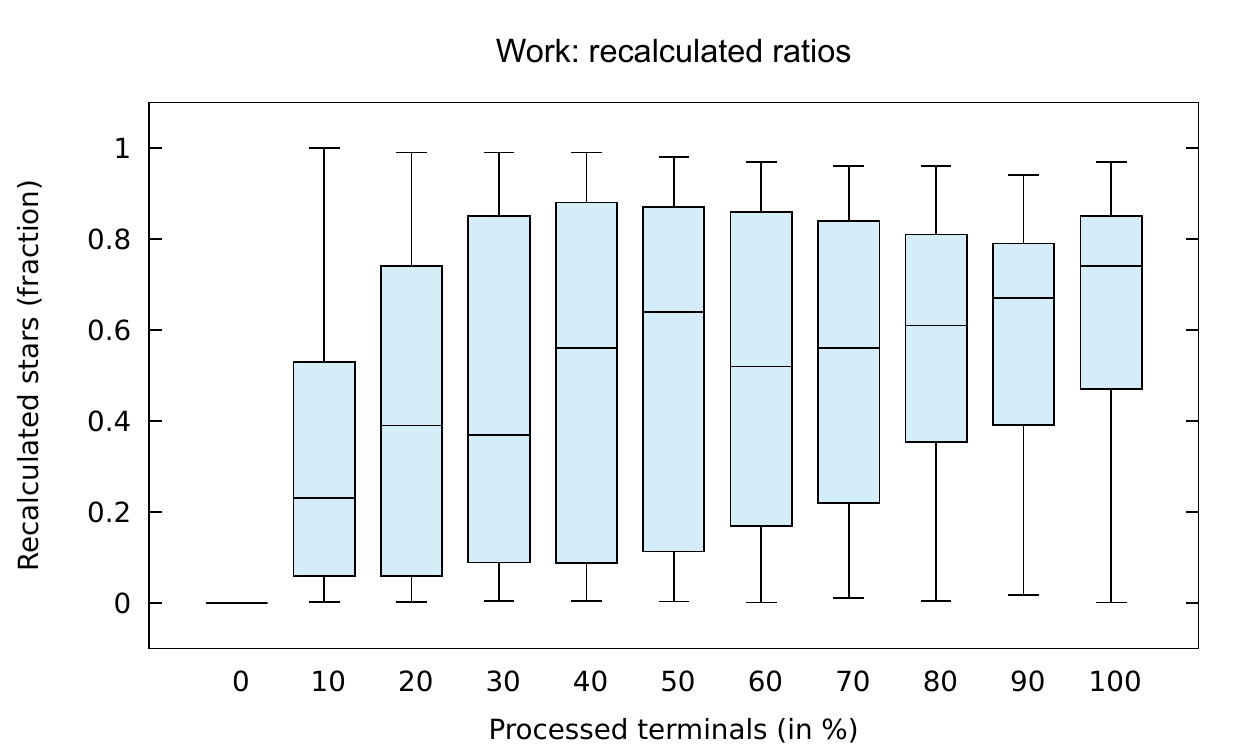}\hfil
  \includegraphics[width=0.48\textwidth]{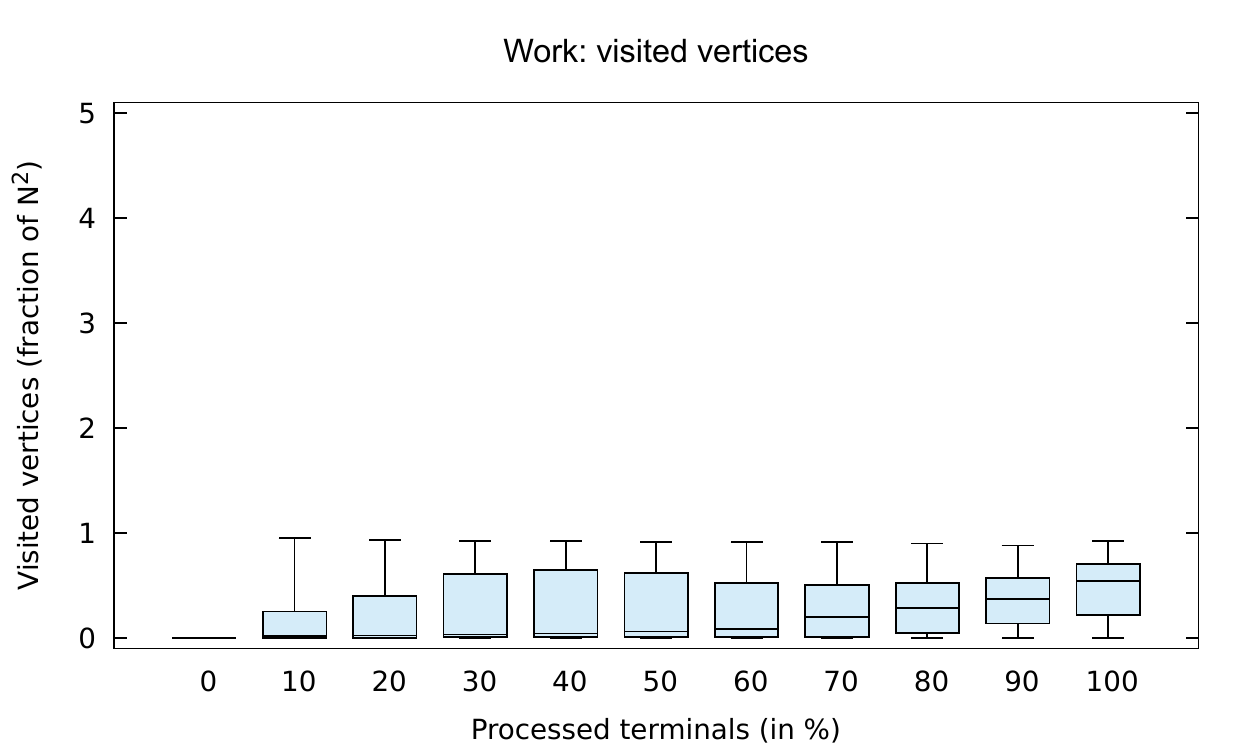}
  \includegraphics[width=0.48\textwidth]{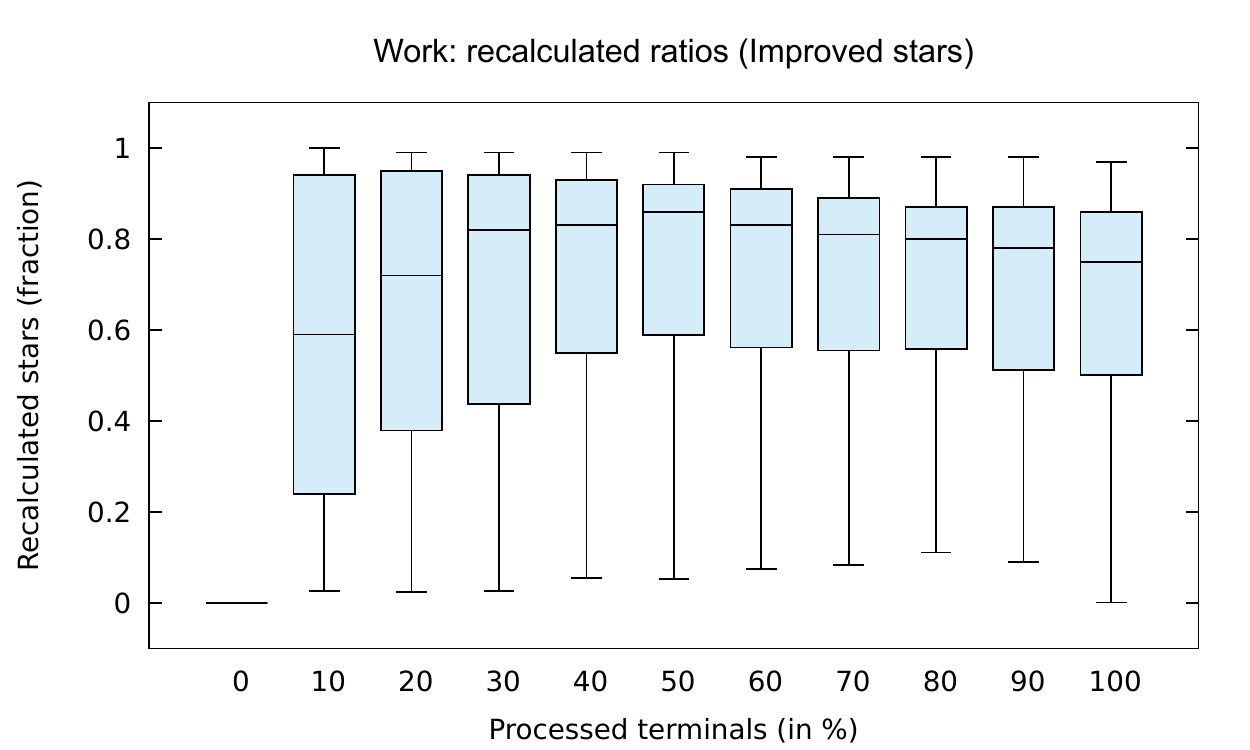}\hfil
  \includegraphics[width=0.48\textwidth]{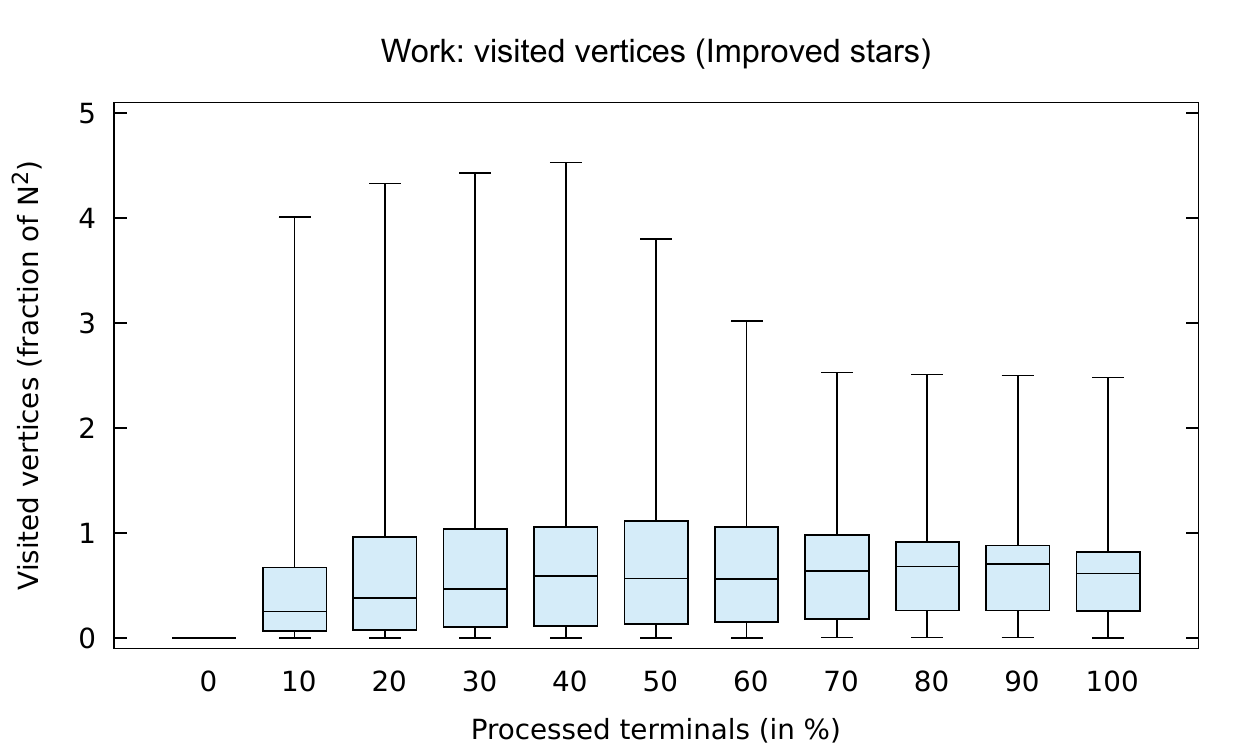}
  \includegraphics[width=0.48\textwidth]{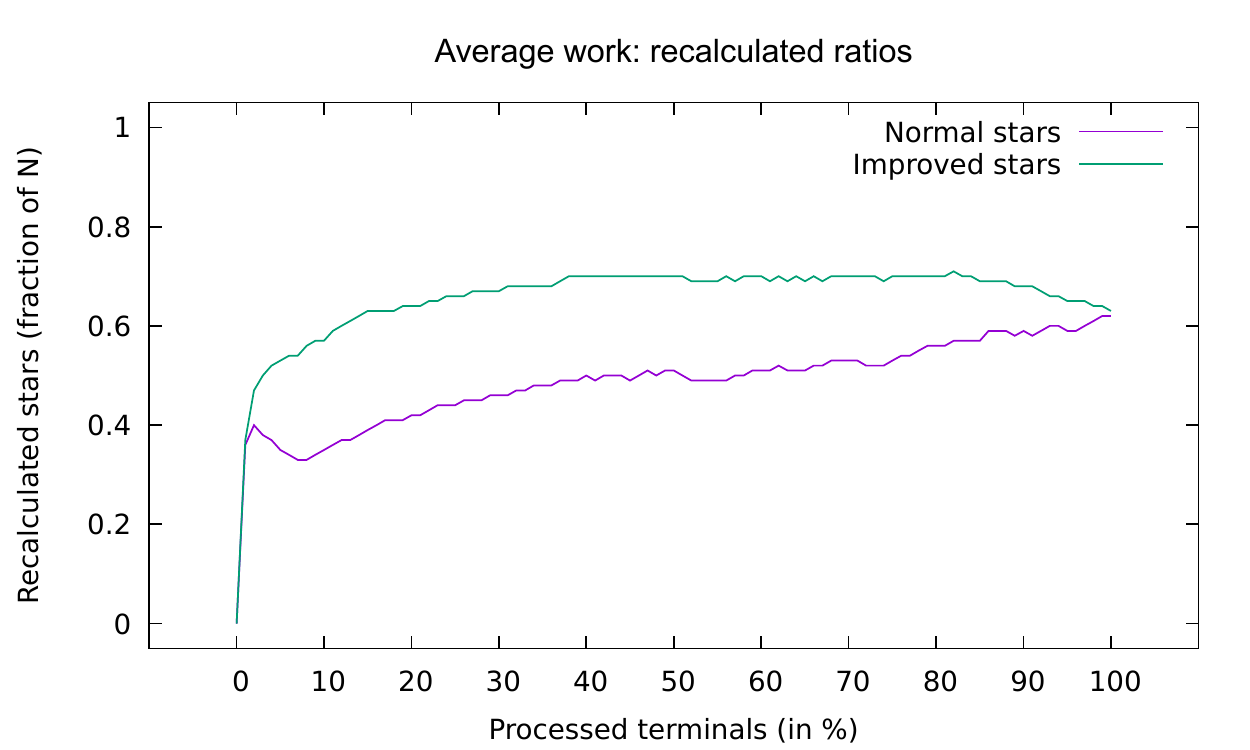}\hfil
  \includegraphics[width=0.48\textwidth]{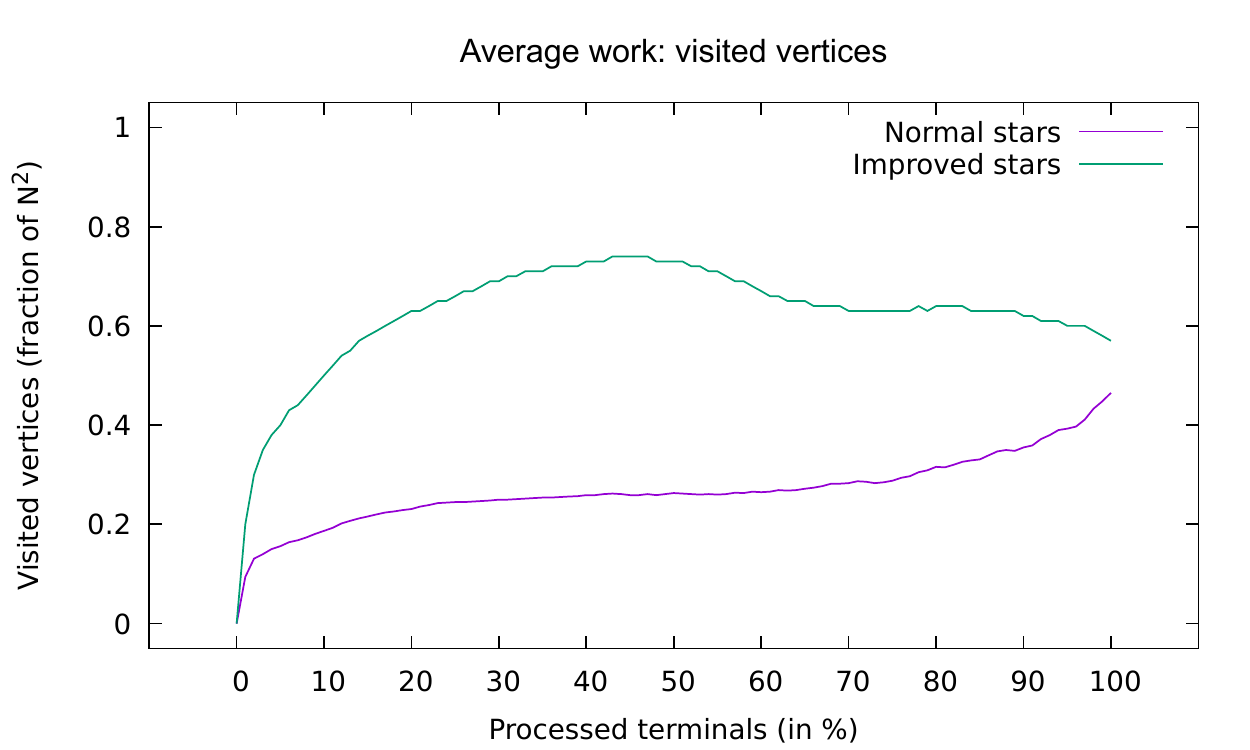}
\medskip
  \caption{\label{fig:work}%
  Work done on PACE Challenge 2018 instances. (For raw data see Table~\ref{tab:work}.)}
\end{figure*}

When comparing improved stars (Section~\ref{sec:findingBestStar}) with basic stars we get a similar performance after the first few contractions.
However, improved stars improve the solution quite importantly, even cooperates well in combination with MST+.
This is far the best method we studied when aiming at the smallest solution as it differs from the ``best'' solution by at most 6.49\% and in the median by only 0.11\%.
The downside is the slower running time.
We describe this difference in Figure~\ref{fig:work}. 
It clearly shows that both the number of recalculated ratios and the number of visited vertices is significantly larger for them. 

Surprisingly Star Contractions improve even the performance of the classical implementation of Zelikovsky's algorithm.
However, this is easily outperformed by improved stars combined with MST+ algorithm.
On the other hand, our variants Zelikovsky$-$ and Zelikovsky+ behaves reasonably well from the beginning and it seems that stars contractions do only a little with it.
For example, Zelikovsky+ has a median that is only 1.04\% worse even without any star contractions compared to 3.59\% (for MST+).
This good performance even without any star contractions is diminished by much slower running time which is comparable with many rounds of star contractions finished by MST+.
Therefore, it is questionable what is more useful in practice (compared to resources of consumption).
However, this was not in the scope of our experiments and so we leave this for further investigations and comparisons.

As we have already observed, if during the algorithm's execution we only contract stars containing only two terminals, then the proposed algorithm returns a minimal spanning tree in the metric closure of the graph on terminals.
Therefore, one should expect that if stars contracting more terminals are found and contracted during the execution, then the quality of the solution found should improve.
We have implemented two possible ways of computing the best star---a basic and an improved one (which tries to bypass the usual overcounting).
Clearly, the later allows us to identify best stars containing more terminals; see Figure~\ref{fig:starSizes}\sv{ in the appendix} and Table~\ref{tab:starSizes}.
Although, as Figure~\ref{fig:aggregatedMST} suggests, the aggregated overall performance is comparable to basic stars, a slightly better solution can be found via improved stars.
Furthermore, we can confirm that in the PACE Challenge 2018 instances:%
\lv{\begin{itemize}}
  \lv{\item}\sv{(1)} the number of stars containing only two terminals decreased by about $8\%$,
  \lv{\item}\sv{(2)} the number of stars with three terminals decreased by $36\%$, while
\lv{\item} \sv{(3)} the number of stars with four to ten terminals increased by $171\%$!
\lv{\end{itemize}}

\toappendix{
\begin{figure*}[bt]
  \centering
   \begin{minipage}[c]{0.78\textwidth}
  \includegraphics[width=\textwidth]{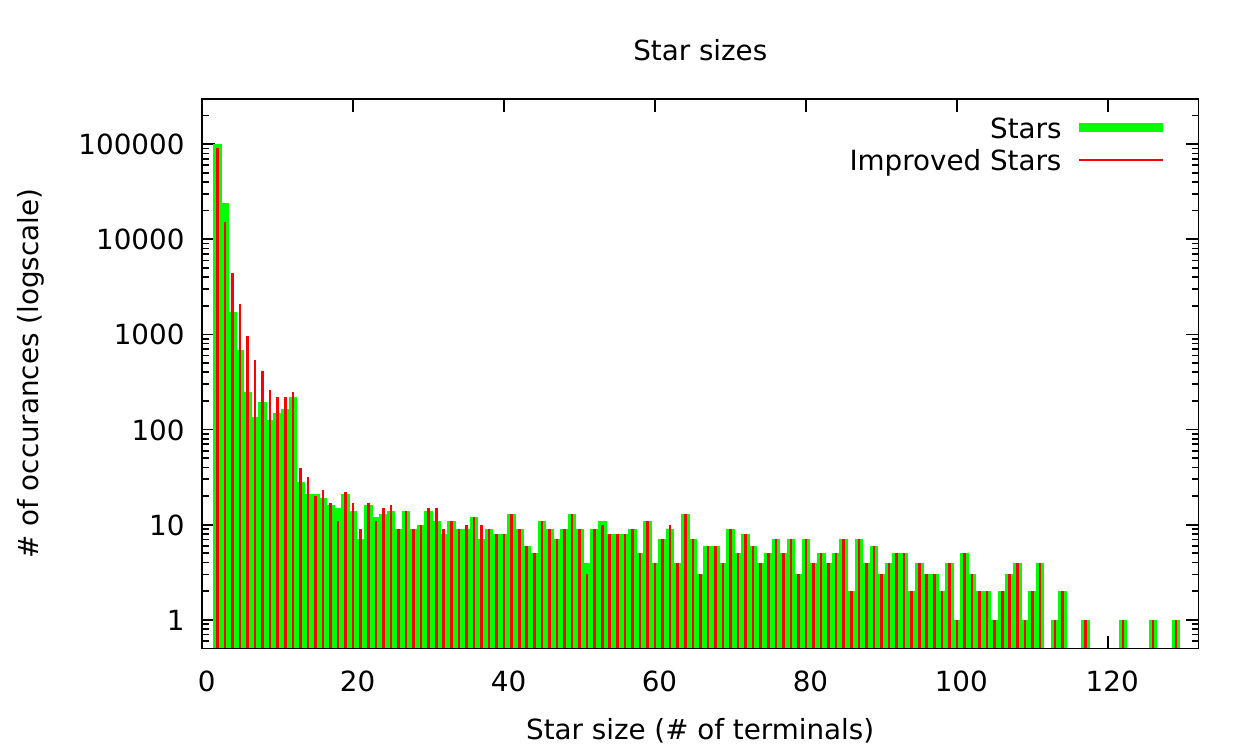}
  \end{minipage}\hfill
  \begin{minipage}[c]{0.21\textwidth}
  \caption{\label{fig:starSizes}%
  Star sizes on PACE Challenge 2018 instances with star contractions running until the end.}
    \end{minipage}
  \end{figure*}
}

\begin{table*}[t]
  \centering
  \include{plots/pace/star-sizes}
\medskip
  \caption{\label{tab:starSizes}%
  The total number of stars containing two to ten terminals contracted during the execution of the algorithm on all PACE Challenge 2018 instances.
  }
\end{table*}

\section{Conclusions}\label{sec:conclusions}
In general, we have confirmed that contracting a best star improves the quality of the solution returned by the MST (MST+) algorithm.
It seems that if we exhaustively apply contractions of best stars, we achieve a solution of slightly better quality than our modification of Zelikovsky's algorithm (i.e., Zelikovsky+ algorithm applied directly to the input).
However running times of such approaches is comparable in practice.
Unfortunately, unlike in classical Zelikovsky's algorithm star contractions do not significantly help in our modifications.
Importantly, MST+ heuristics should replace the classical MST since it outperforms it without being much more complicated or time-consuming.
Improved stars with MST+ do perform better when aiming for the best quality of the solution when the slight increase in the running time is not important.

\smallskip
\noindent{\sffamily\textbf{Solutions with Few Steiner Vertices.}}~~
It is hard to argue about the practicality of this approach to \textsc{Steiner Tree} since it is not clear how to find a good solution to compare with.
Recall that theory suggests~\cite{DvorakFKMTV18} that the solution returned is very close to a solution that is on the one hand cheap (i.e., close to optimum) while on the other hand use few Steiner vertices.
Thus, one question we would like to raise here is how to find such solutions.
In the used data-set from the PACE Challenge 2018, we can identify two extremal groups of instances---first, where our algorithm improves the current solution significantly and the second, where the quality of the solution stays nearly the same throughout the computation.
Can it be that the first set of instances admits a high-quality solution with only a few terminals?
It will be interesting to research whether there are instances from practice with such solutions?

Last but not least, our experiments suggest that our methods for lessening the time needed to compute the best star are useless when there are only a few terminals left in the graph since in such a case the computation is only local.
Yet if this happens (according to Figure~\ref{fig:work} this happens when about 30\% of terminals are left), it is not possible to use the algorithm of Dreyfus and Wagner, since in such cases we still usually have more than 30 terminals left---which is clearly intractable for larger instances.
This brings us to the following question:
Is it possible to use best stars to fasten (in both theory or practice) the algorithm of Dreyfus and Wagner while loosing only a bit in its precision?
If yes, we hope that a suitable combination of the two algorithms can be used in practice.
One can use recent improvements of Dreyfus and Wagner algorithm with the same worst-case running time but which behaves significantly well in practice, in particular on instances originated from VLSI design~\cite{DikstraSteiner}.

\smallskip
\noindent{\sffamily\textbf{Future work.}}~~
A possible future direction includes implementation of more involved approximation algorithms.
An interesting research direction is to augment the algorithm of Dvořák et al.~\cite{DvorakFKMTV18} for Euclidean instances, since solutions to such instances should contain fewer Steiner vertices and thus the quality of the returned solution should increase.
In a similar direction, we performed several basic tests for rectilinear instances
from ORlib (See Section~\ref{sec:rect} in the appendix).
Interestingly, our approach works reasonably well on such specialized instances,
i.e., it performs significantly better than on general instances.
However, we leave this area for future work as the comparison with specialized
heuristics for those instances is essential.

Another direction for future interesting research is a comparison between star contractions and local search or mixed-integer programming techniques.
\lv{Clearly, one can insert more involved approximation algorithms into our framework.}
Yet another possibility is to, instead of greedily contracting a subgraph with
the best ratio, contract a subgraph with a slightly worse ratio which contains substantially many terminals.
As our results suggest the subgraph containing more terminals tends to improve the current as well as the final solution better.
In a broader context, the algorithm of Dvořák et al.~\cite{DvorakFKMTV18} cannot approximate \textsc{Steiner Arborescence} well, since even approximation is hard from parameterized complexity view.
Is this still true in practice?

\lv{%
\smallskip
\noindent{\sffamily\textbf{Acknowledgment.}}~~
We thank the authors of the Boost library~\cite{BoostLib}, which we use in our code, for their work and effective implementation of many graph algorithms.
We thank Tomáš Toufar for consultations in the early stages of preparation of the experiments in the paper as well as for implementation of a part~\cite{HTKME} which was partially used in our code. 
We also thank PACE challenge which was a good motivation and inspiration for developement of practical algorithms for the Steiner Tree problem.
The results of this challenge motivated us to develop the experiments presented in this paper.
Part of the work was carried out while D.~Knop and T.~Masařík were at the University of Bergen.
}

\bibliography{stree}

\clearpage
\appendix

\section{Data used to generate figures in the paper}\label{sec:data}
\toappendix{%
\begin{table*}[htb!]%
  \centering
  \scalebox{0.85}{\begin{tabular}{r|ccccccc}
      \multicolumn{8}{c}{{\bf Averages }} \\
      & MST & MST+ & MST (I) & MST+ (I) & Zel & Zel$-$ & Zel+ \\ \hline\hline
0
& 118.11
& 111.34
& 118.11
& 111.34
& 109.99
& 102.91
& 102.48
\\
10
& 115.64
& 109.65
& 115.67
& 109.73
& 108.80
& 102.83
& 102.41
\\
20
& 113.36
& 108.38
& 113.36
& 108.22
& 107.89
& 102.79
& 102.35
\\
30
& 111.33
& 107.17
& 111.29
& 107.23
& 107.07
& 102.70
& 102.29
\\
40
& 109.59
& 106.46
& 109.40
& 106.30
& 106.41
& 102.47
& 102.14
\\
50
& 108.08
& 105.55
& 107.82
& 105.48
& 105.59
& 102.19
& 101.88
\\
60
& 106.58
& 104.81
& 106.23
& 104.55
& 104.82
& 102.02
& 101.76
\\
70
& 105.04
& 103.97
& 104.69
& 103.68
& 103.73
& 101.85
& 101.59
\\
80
& 103.61
& 103.03
& 103.30
& 102.76
& 102.76
& 101.66
& 101.45
\\
90
& 102.33
& 102.11
& 101.97
& 101.78
& 101.90
& 101.48
& 101.33
\\
100
& 101.36
& 101.24
& 101.02
& 100.94
& 101.36
& 101.36
& 101.24
\\
\end{tabular}
}
\medskip
  \caption{\label{tab:aggregatedMST2}%
  Data used to generate Figure~\ref{fig:aggregatedMST} part II.}
\end{table*}

\begin{table*}[tb]%
  \centering
  \scalebox{0.85}{\begin{tabular}{r|ccccc}
  \multicolumn{6}{c}{{\bf MST}} \\
  & Min & 1st q. & Median & 3rd q. & Max \\ \hline\hline
  0 & 100.00 & 100.31 & 110.01 & 134.26 & 169.07
\\ 10 & 100.00 & 100.31 & 109.97 & 129.91 & 162.67
\\ 20 & 100.00 & 100.30 & 109.28 & 123.14 & 156.26
\\ 30 & 100.00 & 100.29 & 108.69 & 117.30 & 149.85
\\ 40 & 100.00 & 100.27 & 107.70 & 113.95 & 143.02
\\ 50 & 100.00 & 100.26 & 106.26 & 111.97 & 135.33
\\ 60 & 100.00 & 100.22 & 104.96 & 109.76 & 129.42
\\ 70 & 100.00 & 100.15 & 104.01 & 107.59 & 120.53
\\ 80 & 100.00 & 100.13 & 103.28 & 105.49 & 114.00
\\ 90 & 100.00 & 100.12 & 102.20 & 103.99 & 111.11
\\ 100 & 100.00 & 100.01 & 100.51 & 102.17 & 111.11
\end{tabular}
}\hfil
  \scalebox{0.85}{\begin{tabular}{r|ccccc}
  \multicolumn{6}{c}{{\bf MST (Improved stars)}} \\
  & Min & 1st q. & Median & 3rd q. & Max \\ \hline\hline
  0 & 100.00 & 100.31 & 110.01 & 134.26 & 169.07
\\ 10 & 100.00 & 100.31 & 109.97 & 130.07 & 161.70
\\ 20 & 100.00 & 100.30 & 109.25 & 123.24 & 154.32
\\ 30 & 100.00 & 100.29 & 108.66 & 116.92 & 147.18
\\ 40 & 100.00 & 100.27 & 107.71 & 113.43 & 139.70
\\ 50 & 100.00 & 100.25 & 106.03 & 111.54 & 132.73
\\ 60 & 100.00 & 100.22 & 104.71 & 108.74 & 129.42
\\ 70 & 100.00 & 100.15 & 103.97 & 106.66 & 120.53
\\ 80 & 100.00 & 100.12 & 102.91 & 104.81 & 114.00
\\ 90 & 100.00 & 100.12 & 101.75 & 103.22 & 106.61
\\ 100 & 100.00 & 100.01 & 100.51 & 101.74 & 106.61
\end{tabular}
} \medskip\\
  \scalebox{0.85}{\begin{tabular}{r|ccccc}
  \multicolumn{6}{c}{{\bf MST+}} \\
  & Min & 1st q. & Median & 3rd q. & Max \\ \hline\hline
  0 & 100.00 & 100.24 & 107.39 & 116.69 & 146.66
\\ 10 & 100.00 & 100.24 & 107.33 & 113.92 & 143.23
\\ 20 & 100.00 & 100.24 & 106.93 & 113.05 & 139.37
\\ 30 & 100.00 & 100.23 & 106.14 & 109.86 & 133.47
\\ 40 & 100.00 & 100.23 & 105.82 & 108.81 & 130.83
\\ 50 & 100.00 & 100.22 & 104.63 & 108.09 & 128.88
\\ 60 & 100.00 & 100.17 & 103.92 & 107.05 & 126.93
\\ 70 & 100.00 & 100.13 & 103.11 & 105.96 & 120.02
\\ 80 & 100.00 & 100.11 & 102.70 & 104.75 & 112.43
\\ 90 & 100.00 & 100.11 & 101.79 & 103.13 & 111.11
\\ 100 & 100.00 & 100.00 & 100.18 & 102.00 & 111.11
\end{tabular}
}\hfil
  \scalebox{0.85}{\begin{tabular}{r|ccccc}
  \multicolumn{6}{c}{{\bf MST+ (Improved stars)}} \\
  & Min & 1st q. & Median & 3rd q. & Max \\ \hline\hline
  0 & 100.00 & 100.24 & 107.39 & 116.69 & 146.66
\\ 10 & 100.00 & 100.24 & 107.39 & 115.38 & 143.23
\\ 20 & 100.00 & 100.24 & 106.97 & 112.23 & 139.37
\\ 30 & 100.00 & 100.23 & 106.08 & 110.09 & 133.47
\\ 40 & 100.00 & 100.23 & 105.82 & 108.86 & 128.39
\\ 50 & 100.00 & 100.22 & 104.36 & 108.56 & 127.52
\\ 60 & 100.00 & 100.17 & 103.67 & 106.80 & 122.38
\\ 70 & 100.00 & 100.13 & 103.07 & 105.62 & 119.46
\\ 80 & 100.00 & 100.11 & 102.28 & 104.30 & 112.31
\\ 90 & 100.00 & 100.11 & 101.58 & 102.98 & 106.49
\\ 100 & 100.00 & 100.00 & 100.20 & 101.63 & 106.49
\end{tabular}
} \medskip\\
  \scalebox{0.85}{\begin{tabular}{r|ccccc}
  \multicolumn{6}{c}{{\bf Zelikovsky}} \\
  & Min & 1st q. & Median & 3rd q. & Max \\ \hline\hline
  0 & 100.00 & 100.15 & 103.45 & 114.94 & 165.01
\\ 10 & 100.00 & 100.16 & 104.11 & 112.45 & 158.90
\\ 20 & 100.00 & 100.16 & 104.05 & 112.17 & 152.02
\\ 30 & 100.00 & 100.16 & 104.09 & 111.92 & 145.26
\\ 40 & 100.00 & 100.15 & 103.52 & 109.96 & 137.38
\\ 50 & 100.00 & 100.14 & 103.44 & 108.27 & 130.30
\\ 60 & 100.00 & 100.14 & 103.01 & 106.59 & 126.56
\\ 70 & 100.00 & 100.12 & 102.46 & 105.36 & 119.16
\\ 80 & 100.00 & 100.11 & 102.19 & 104.01 & 112.25
\\ 90 & 100.00 & 100.11 & 101.72 & 102.64 & 111.11
\\ 100 & 100.00 & 100.01 & 100.51 & 102.17 & 111.11
\end{tabular}
}\hfil
  \scalebox{0.85}{\begin{tabular}{r|ccccc}
  \multicolumn{6}{c}{{\bf Zelikovsky-}} \\
  & Min & 1st q. & Median & 3rd q. & Max \\ \hline\hline
  0 & 100.00 & 100.07 & 101.42 & 103.73 & 118.20
\\ 10 & 100.00 & 100.07 & 101.42 & 103.58 & 118.20
\\ 20 & 100.00 & 100.07 & 101.33 & 103.67 & 118.20
\\ 30 & 100.00 & 100.08 & 101.32 & 103.52 & 118.20
\\ 40 & 100.00 & 100.08 & 101.38 & 103.24 & 117.23
\\ 50 & 100.00 & 100.08 & 101.35 & 103.18 & 113.90
\\ 60 & 100.00 & 100.08 & 101.23 & 103.12 & 111.44
\\ 70 & 100.00 & 100.07 & 101.22 & 102.84 & 111.11
\\ 80 & 100.00 & 100.06 & 101.33 & 102.40 & 111.11
\\ 90 & 100.00 & 100.07 & 101.06 & 102.19 & 111.11
\\ 100 & 100.00 & 100.01 & 100.51 & 102.17 & 111.11
\end{tabular}
} \medskip\\
  \scalebox{0.85}{\begin{tabular}{r|ccccc}
  \multicolumn{6}{c}{{\bf Zelikovsky+}} \\
  & Min & 1st q. & Median & 3rd q. & Max \\ \hline\hline
  0 & 100.00 & 100.06 & 101.42 & 103.20 & 116.16
\\ 10 & 100.00 & 100.06 & 101.42 & 103.10 & 115.15
\\ 20 & 100.00 & 100.07 & 101.33 & 103.38 & 115.15
\\ 30 & 100.00 & 100.07 & 101.24 & 103.17 & 115.15
\\ 40 & 100.00 & 100.07 & 101.31 & 103.06 & 114.23
\\ 50 & 100.00 & 100.07 & 101.21 & 102.67 & 112.55
\\ 60 & 100.00 & 100.07 & 101.07 & 102.51 & 111.11
\\ 70 & 100.00 & 100.06 & 101.03 & 102.35 & 111.11
\\ 80 & 100.00 & 100.06 & 100.99 & 102.15 & 111.11
\\ 90 & 100.00 & 100.07 & 100.59 & 102.03 & 111.11
\\ 100 & 100.00 & 100.00 & 100.18 & 102.00 & 111.11
\end{tabular}
}
\medskip
  \caption{\label{tab:aggregatedMST}%
  Data used to generate Figure~\ref{fig:aggregatedMST} part I.}
\end{table*}
}
\toappendix{%
\begin{table*}
  \centering
  \scalebox{0.85}{\begin{tabular}{r|ccccc}
  \multicolumn{6}{c}{{\bf Work: recalculated ratios}} \\
  & Min & 1st q. & Median & 3rd q. & Max \\ \hline\hline
  0 & 0.00000 & 0.000 & 0.00 & 0.00 & 0.00
\\ 10 & 0.00275 & 0.060 & 0.23 & 0.53 & 1.00
\\ 20 & 0.00135 & 0.060 & 0.39 & 0.74 & 0.99
\\ 30 & 0.00465 & 0.089 & 0.37 & 0.85 & 0.99
\\ 40 & 0.00448 & 0.088 & 0.56 & 0.88 & 0.99
\\ 50 & 0.00312 & 0.113 & 0.64 & 0.87 & 0.98
\\ 60 & 0.00058 & 0.169 & 0.52 & 0.86 & 0.97
\\ 70 & 0.01095 & 0.220 & 0.56 & 0.84 & 0.96
\\ 80 & 0.00418 & 0.353 & 0.61 & 0.81 & 0.96
\\ 90 & 0.01804 & 0.391 & 0.67 & 0.79 & 0.94
\\ 100 & 0.00064 & 0.470 & 0.74 & 0.85 & 0.97
\end{tabular}
}\hfil
  \scalebox{0.85}{\begin{tabular}{r|ccccc}
  \multicolumn{6}{c}{{\bf Work: visited vertices}} \\
  & Min & 1st q. & Median & 3rd q. & Max \\ \hline\hline
  0 & 0.0e+00 & 0.0000 & 0.000 & 0.000 & 0.00
\\ 10 & 8.7e-06 & 0.0019 & 0.017 & 0.250 & 0.95
\\ 20 & 6.6e-06 & 0.0021 & 0.023 & 0.402 & 0.93
\\ 30 & 2.7e-05 & 0.0052 & 0.032 & 0.606 & 0.92
\\ 40 & 2.7e-05 & 0.0054 & 0.043 & 0.644 & 0.92
\\ 50 & 1.3e-05 & 0.0060 & 0.060 & 0.616 & 0.91
\\ 60 & 7.4e-06 & 0.0110 & 0.086 & 0.527 & 0.91
\\ 70 & 1.3e-04 & 0.0132 & 0.201 & 0.505 & 0.91
\\ 80 & 3.7e-05 & 0.0469 & 0.287 & 0.520 & 0.90
\\ 90 & 2.1e-04 & 0.1362 & 0.370 & 0.573 & 0.88
\\ 100 & 6.1e-07 & 0.2211 & 0.544 & 0.706 & 0.92
\end{tabular}
}\medskip\\
  \scalebox{0.85}{\begin{tabular}{r|ccccc}
  \multicolumn{6}{c}{{\bf Work: recalculated ratios (Improved stars)}} \\
  & Min & 1st q. & Median & 3rd q. & Max \\ \hline\hline
  0 & 0.00000 & 0.000 & 0.00 & 0.00 & 0.00
\\ 10 & 0.02601 & 0.239 & 0.59 & 0.94 & 1.00
\\ 20 & 0.02415 & 0.379 & 0.72 & 0.95 & 0.99
\\ 30 & 0.02587 & 0.437 & 0.82 & 0.94 & 0.99
\\ 40 & 0.05497 & 0.550 & 0.83 & 0.93 & 0.99
\\ 50 & 0.05177 & 0.589 & 0.86 & 0.92 & 0.99
\\ 60 & 0.07513 & 0.562 & 0.83 & 0.91 & 0.98
\\ 70 & 0.08319 & 0.555 & 0.81 & 0.89 & 0.98
\\ 80 & 0.11053 & 0.558 & 0.80 & 0.87 & 0.98
\\ 90 & 0.09016 & 0.512 & 0.78 & 0.87 & 0.98
\\ 100 & 0.00064 & 0.502 & 0.75 & 0.86 & 0.97
\end{tabular}
}\hfil
  \scalebox{0.85}{\begin{tabular}{r|ccccc}
  \multicolumn{6}{c}{{\bf Work: visited vertices (Improved stars)}} \\
  & Min & 1st q. & Median & 3rd q. & Max \\ \hline\hline
  0 & 0.0e+00 & 0.000 & 0.000 & 0.00 & 0.00
\\ 10 & 5.7e-04 & 0.066 & 0.252 & 0.67 & 4.01
\\ 20 & 6.1e-04 & 0.079 & 0.381 & 0.96 & 4.33
\\ 30 & 6.8e-04 & 0.106 & 0.468 & 1.04 & 4.43
\\ 40 & 1.8e-03 & 0.112 & 0.588 & 1.06 & 4.53
\\ 50 & 1.5e-03 & 0.133 & 0.567 & 1.11 & 3.80
\\ 60 & 2.7e-03 & 0.150 & 0.564 & 1.06 & 3.02
\\ 70 & 3.7e-03 & 0.181 & 0.641 & 0.98 & 2.53
\\ 80 & 4.9e-03 & 0.263 & 0.681 & 0.91 & 2.51
\\ 90 & 4.0e-03 & 0.261 & 0.703 & 0.88 & 2.50
\\ 100 & 6.1e-07 & 0.254 & 0.615 & 0.82 & 2.48
\end{tabular}
} \medskip\\
  \scalebox{0.85}{\begin{tabular}{r|cc}
      \multicolumn{3}{c}{{\bf Averages: recalculated ratios }} \\
      & Normal & Improved \\ \hline\hline
0
& 0.00
& 0.00
\\
10
& 0.57
& 0.35
\\
20
& 0.64
& 0.42
\\
30
& 0.67
& 0.46
\\
40
& 0.70
& 0.50
\\
50
& 0.70
& 0.51
\\
60
& 0.70
& 0.51
\\
70
& 0.70
& 0.53
\\
80
& 0.70
& 0.56
\\
90
& 0.68
& 0.59
\\
100
& 0.63
& 0.62
\\
\end{tabular}
}\hfil
  \scalebox{0.85}{\begin{tabular}{r|cc}
      \multicolumn{3}{c}{{\bf Averages: visited vertices }} \\
      & Normal & Improved \\ \hline\hline
0
& 0.00
& 0.000
\\
10
& 0.50
& 0.187
\\
20
& 0.63
& 0.231
\\
30
& 0.69
& 0.250
\\
40
& 0.73
& 0.259
\\
50
& 0.73
& 0.263
\\
60
& 0.67
& 0.265
\\
70
& 0.63
& 0.283
\\
80
& 0.64
& 0.316
\\
90
& 0.62
& 0.355
\\
100
& 0.57
& 0.465
\\
\end{tabular}
}
  \caption{\label{tab:work}%
  Data used to generate Figure~\ref{fig:work}.}
\end{table*}
}

\appendixText

\clearpage
\newpage
\section{Rectilinear Instances from ORlib}\label{sec:rect}
\begin{figure*}[b]
  \centering
  \includegraphics[width=0.48\textwidth]{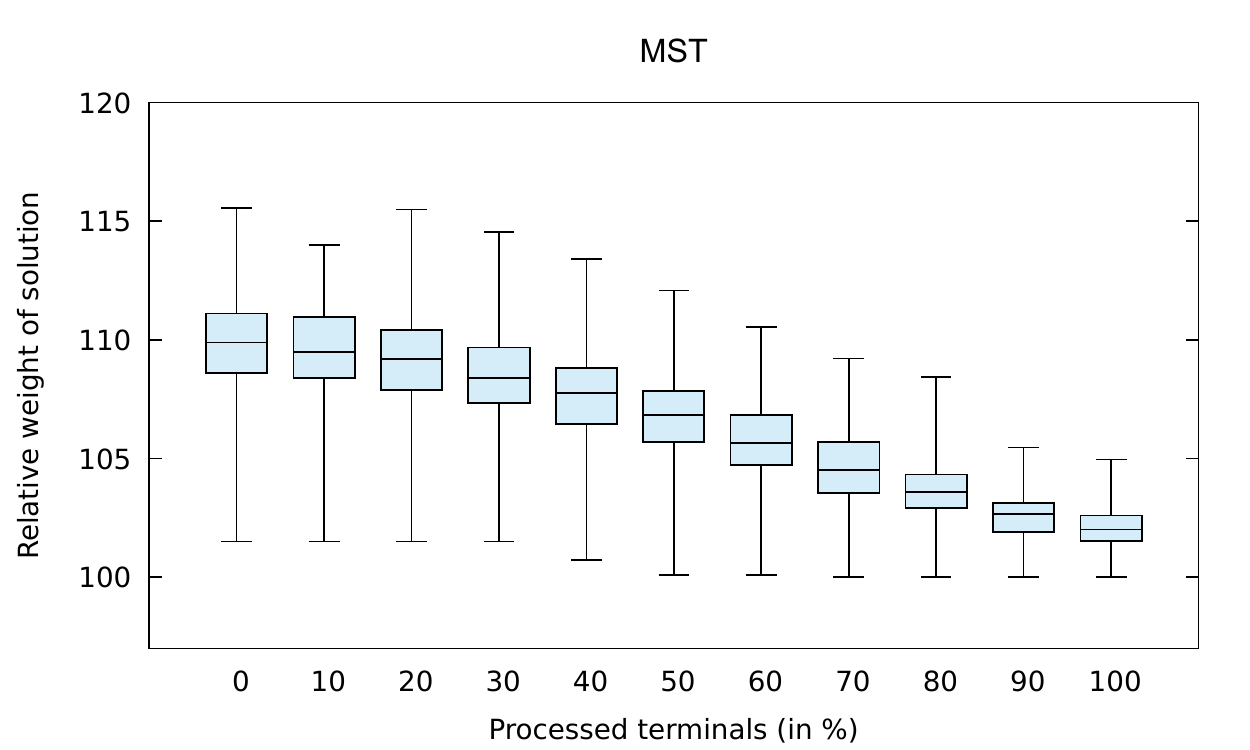}\hfil
  \includegraphics[width=0.48\textwidth]{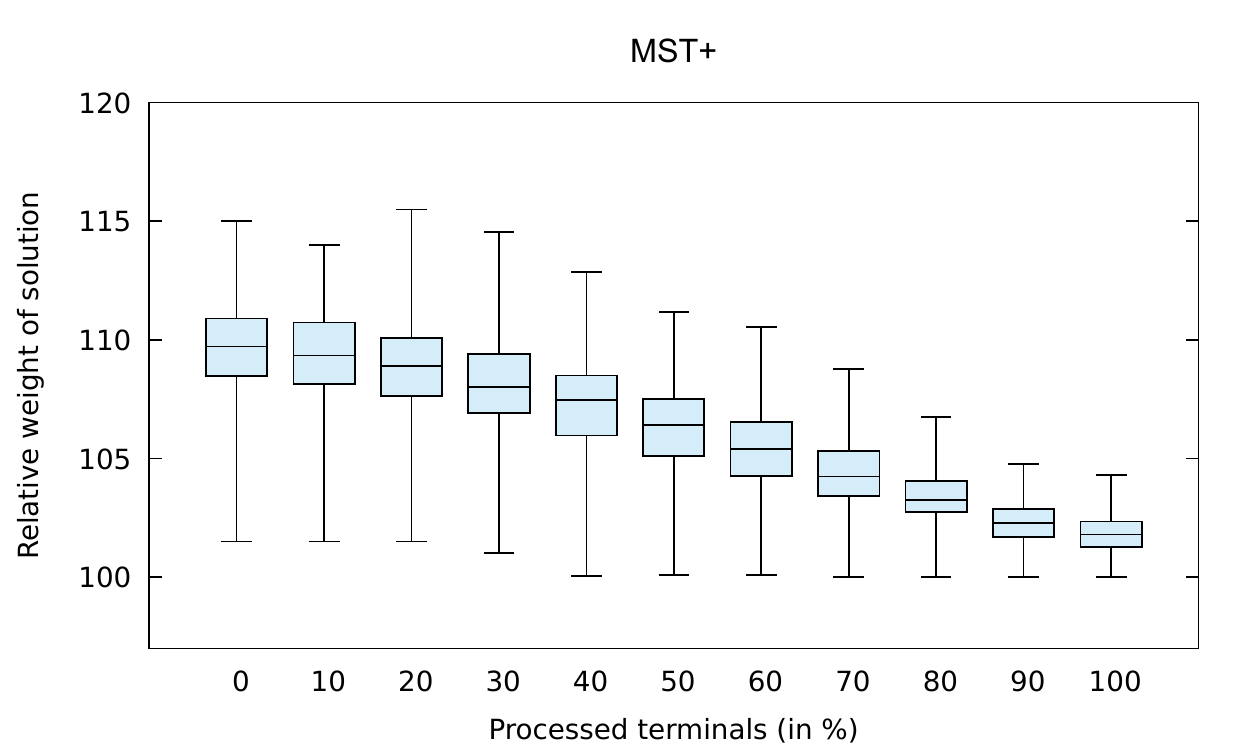}
  \includegraphics[width=0.48\textwidth]{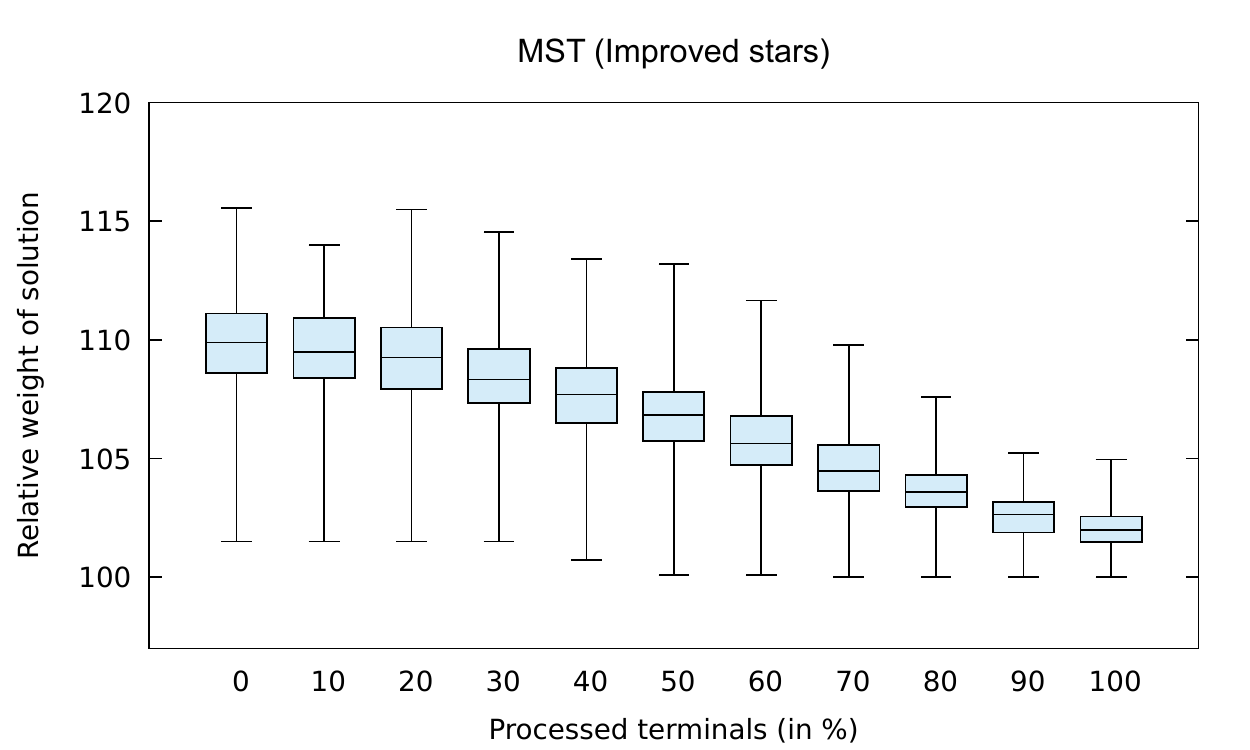}\hfil
  \includegraphics[width=0.48\textwidth]{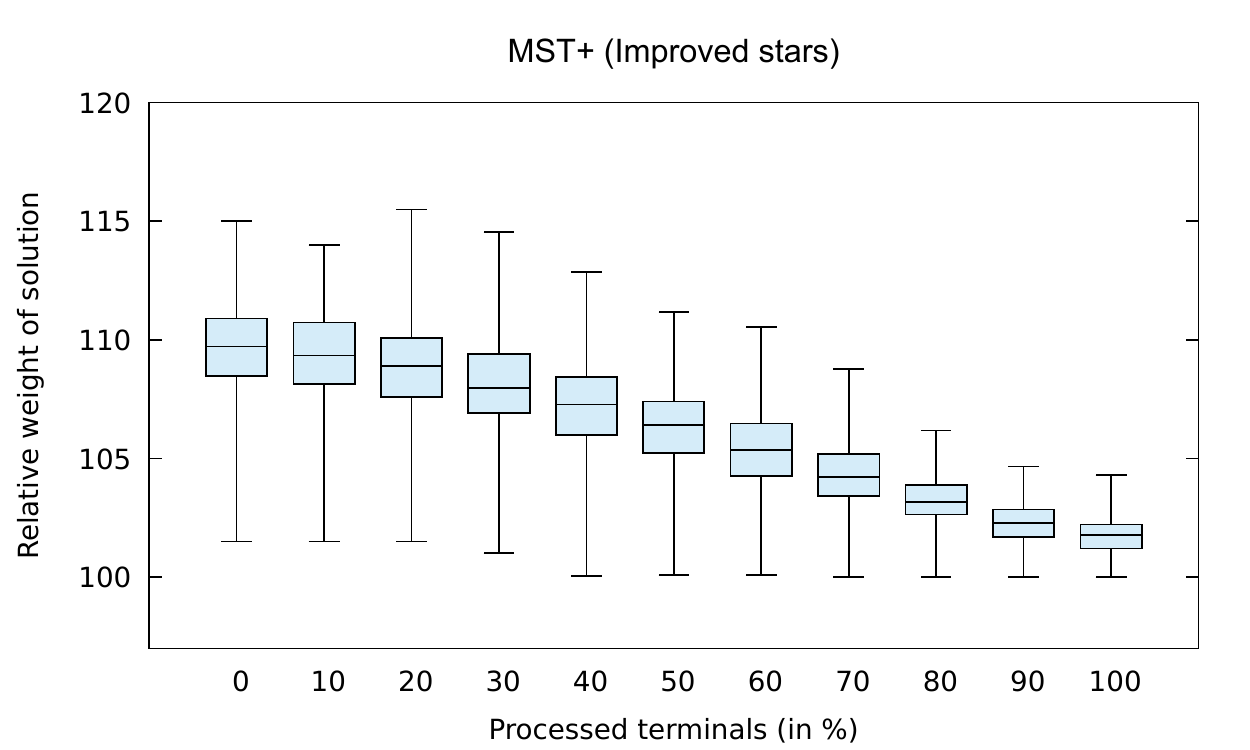}
  \includegraphics[width=0.48\textwidth]{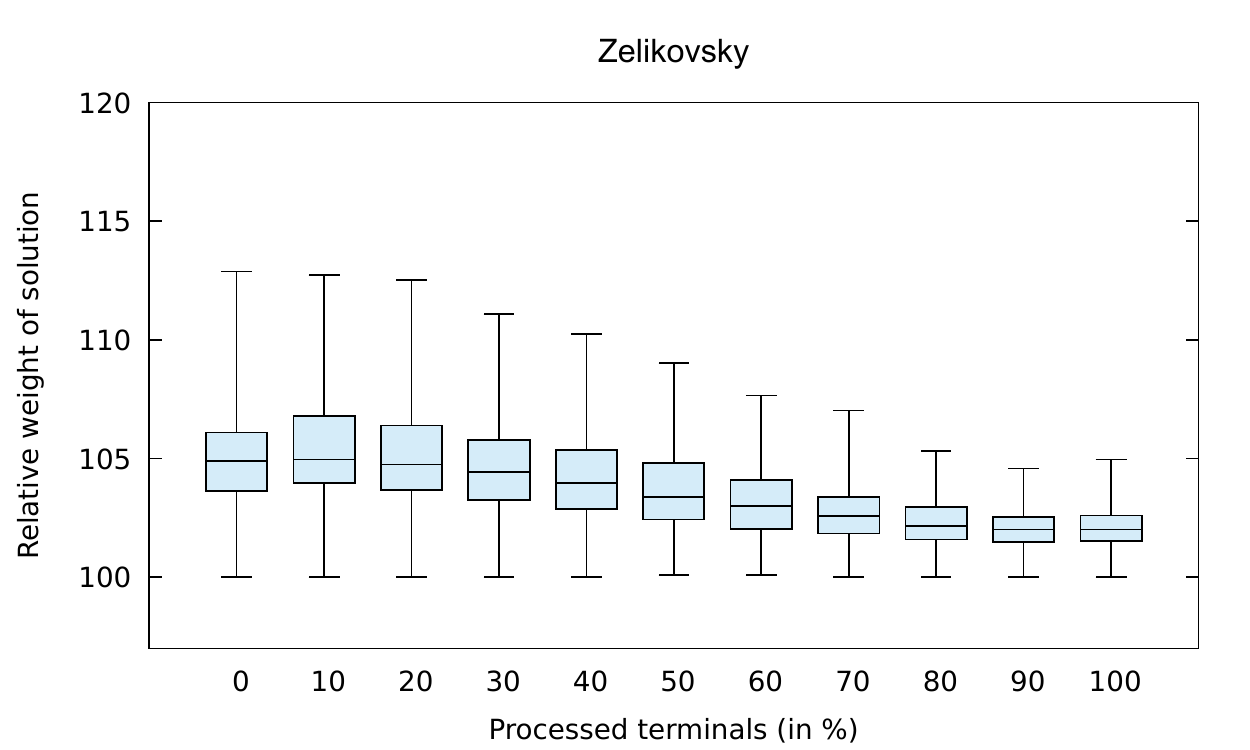}\hfil
  \includegraphics[width=0.48\textwidth]{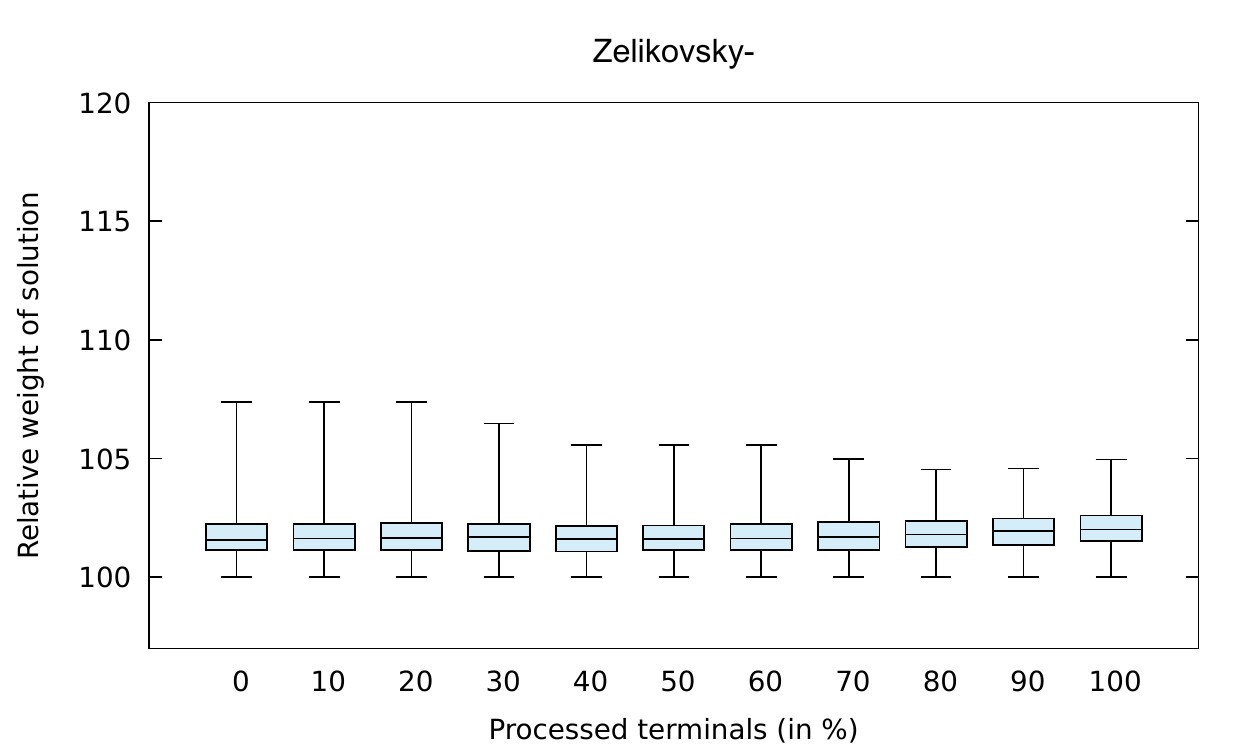}
  \includegraphics[width=0.48\textwidth]{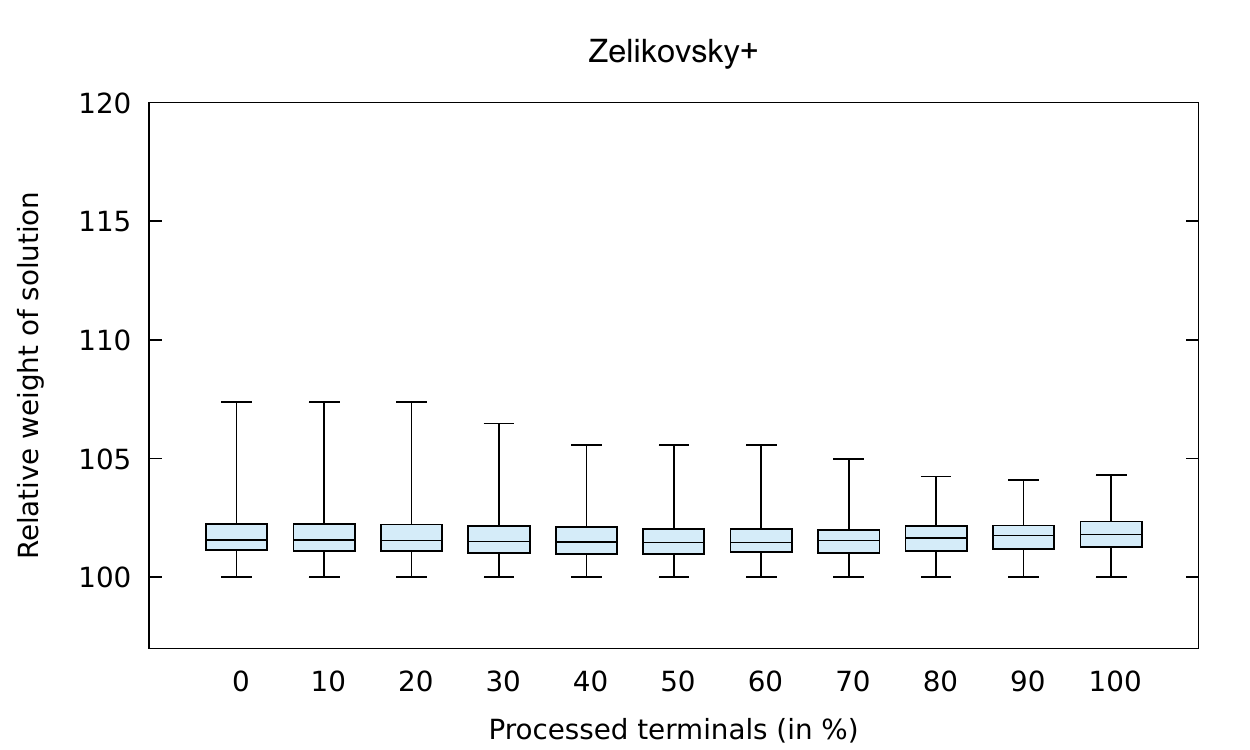}\hfil
  \includegraphics[width=0.48\textwidth]{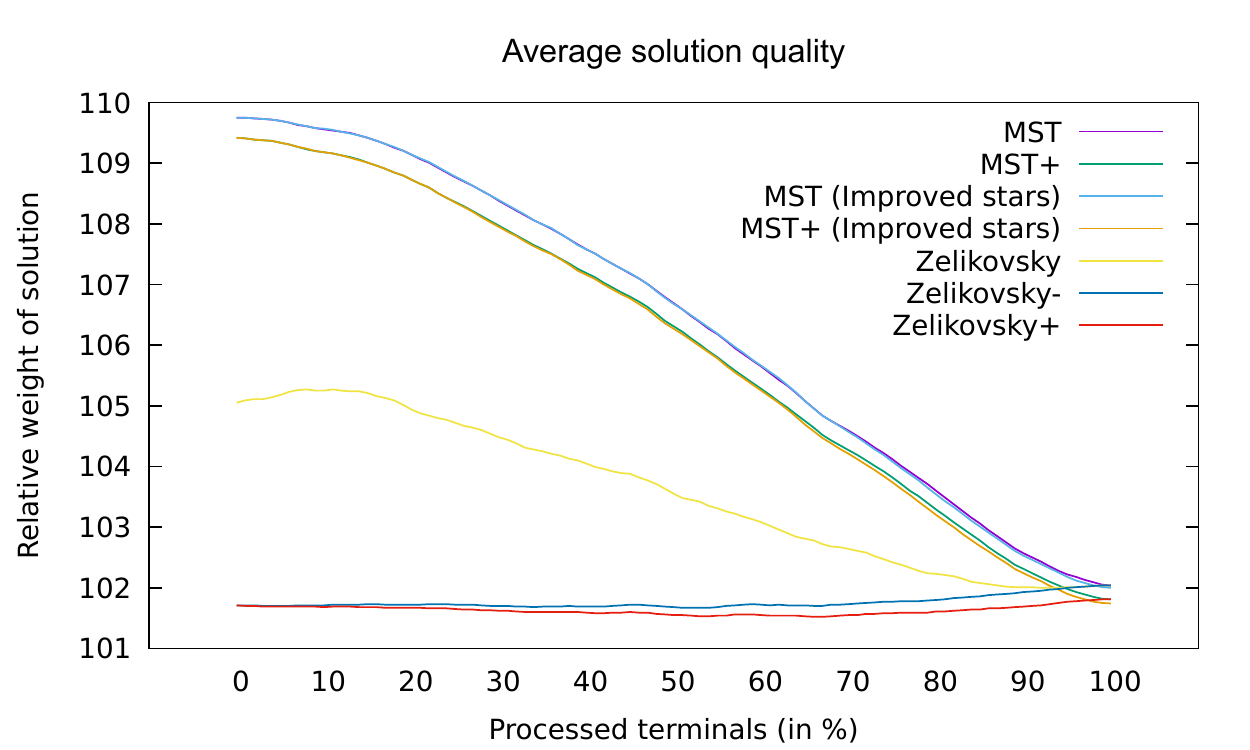}
\medskip
  \caption{\label{fig:Rect}%
  Performance comparison of star contractions and MST heuristic
  (like Figure~\ref{fig:aggregatedMST}) on rectilinear instances obtained from ORLib
  (for raw data see Tables~\ref{tab:Rect} and~\ref{tab:Rect2}).}
\end{figure*}

The same tests as in Section~\ref{sec:PACEInstances} were performed on rectilinear instances from ORlib.
Technically we took data from ORlib rectilinear instances where only positions of Terminals are given in the Euclidean plane.
First, we generated all possible Steiner vertices and therefore our generated instance has quadratic size.

The results suggest that our approach performs well even on rectilinear instances, see Figure~\ref{fig:Rect} and corresponding data in Table~\ref{tab:Rect}.
In fact, the performance on these specialized instances is much better than in the general case.
Here not only some instances improve significantly with star contractions but also the majority of instances is reasonably improved.
Overall, the trends are very similar to the analyses in Section~\ref{sec:PACEInstances}, therefore we will not repeat it here.

As is claimed in the conclusions section this direction is left for further investigation since comparison with methods for rectilinear instances is needed.
\begin{table*}[h!]%
  \centering
  \scalebox{0.87}{\begin{tabular}{r|ccccccc}
      \multicolumn{8}{c}{{\bf Averages }} \\
      & MST & MST+ & MST (I) & MST+ (I) & Zel & Zel$-$ & Zel+ \\ \hline\hline
0
& 109.75
& 109.42
& 109.75
& 109.42
& 105.05
& 101.71
& 101.71
\\
10
& 109.56
& 109.18
& 109.57
& 109.18
& 105.25
& 101.71
& 101.68
\\
20
& 109.14
& 108.73
& 109.14
& 108.73
& 104.94
& 101.72
& 101.67
\\
30
& 108.38
& 107.97
& 108.39
& 107.95
& 104.48
& 101.70
& 101.62
\\
40
& 107.58
& 107.19
& 107.58
& 107.16
& 104.05
& 101.69
& 101.59
\\
50
& 106.69
& 106.31
& 106.68
& 106.27
& 103.55
& 101.68
& 101.55
\\
60
& 105.65
& 105.28
& 105.66
& 105.25
& 103.08
& 101.72
& 101.55
\\
70
& 104.59
& 104.27
& 104.57
& 104.21
& 102.64
& 101.73
& 101.55
\\
80
& 103.60
& 103.29
& 103.54
& 103.20
& 102.23
& 101.80
& 101.61
\\
90
& 102.57
& 102.31
& 102.53
& 102.24
& 102.01
& 101.93
& 101.69
\\
100
& 102.04
& 101.81
& 102.00
& 101.74
& 102.04
& 102.04
& 101.81
\\
\end{tabular}
}
\medskip
  \caption{\label{tab:Rect2}%
  Data used to generate Figure~\ref{fig:Rect} part II.}
\end{table*}

\begin{table*}[h!]%
  \centering
  \scalebox{0.85}{\begin{tabular}{r|ccccc}
  \multicolumn{6}{c}{{\bf MST}} \\
  & Min & 1st q. & Median & 3rd q. & Max \\ \hline\hline
  0 & 101.51 & 108.60 & 109.89 & 111.11 & 115.56
\\ 10 & 101.51 & 108.38 & 109.48 & 110.95 & 114.01
\\ 20 & 101.51 & 107.90 & 109.20 & 110.43 & 115.50
\\ 30 & 101.51 & 107.33 & 108.39 & 109.68 & 114.56
\\ 40 & 100.74 & 106.46 & 107.76 & 108.81 & 113.41
\\ 50 & 100.10 & 105.69 & 106.85 & 107.84 & 112.08
\\ 60 & 100.10 & 104.74 & 105.66 & 106.84 & 110.55
\\ 70 & 100.00 & 103.56 & 104.52 & 105.70 & 109.22
\\ 80 & 100.00 & 102.93 & 103.60 & 104.33 & 108.45
\\ 90 & 100.00 & 101.91 & 102.66 & 103.13 & 105.47
\\ 100 & 100.00 & 101.52 & 102.01 & 102.60 & 104.97
\end{tabular}
}\hfil
  \scalebox{0.85}{\begin{tabular}{r|ccccc}
  \multicolumn{6}{c}{{\bf MST (Improved stars)}} \\
  & Min & 1st q. & Median & 3rd q. & Max \\ \hline\hline
  0 & 101.51 & 108.60 & 109.89 & 111.11 & 115.56
\\ 10 & 101.51 & 108.38 & 109.48 & 110.92 & 114.01
\\ 20 & 101.51 & 107.94 & 109.26 & 110.53 & 115.50
\\ 30 & 101.51 & 107.33 & 108.33 & 109.63 & 114.56
\\ 40 & 100.74 & 106.51 & 107.70 & 108.81 & 113.41
\\ 50 & 100.10 & 105.74 & 106.85 & 107.79 & 113.21
\\ 60 & 100.10 & 104.72 & 105.63 & 106.80 & 111.66
\\ 70 & 100.00 & 103.62 & 104.48 & 105.57 & 109.78
\\ 80 & 100.00 & 102.95 & 103.58 & 104.31 & 107.61
\\ 90 & 100.00 & 101.89 & 102.64 & 103.16 & 105.23
\\ 100 & 100.00 & 101.47 & 102.00 & 102.56 & 104.97
\end{tabular}
} \medskip\\
  \scalebox{0.85}{\begin{tabular}{r|ccccc}
  \multicolumn{6}{c}{{\bf MST+}} \\
  & Min & 1st q. & Median & 3rd q. & Max \\ \hline\hline
  0 & 101.51 & 108.48 & 109.72 & 110.91 & 115.01
\\ 10 & 101.51 & 108.15 & 109.34 & 110.74 & 114.01
\\ 20 & 101.51 & 107.65 & 108.91 & 110.09 & 115.50
\\ 30 & 101.01 & 106.92 & 108.03 & 109.42 & 114.56
\\ 40 & 100.06 & 105.97 & 107.48 & 108.50 & 112.86
\\ 50 & 100.10 & 105.11 & 106.41 & 107.50 & 111.18
\\ 60 & 100.10 & 104.26 & 105.40 & 106.55 & 110.55
\\ 70 & 100.00 & 103.41 & 104.25 & 105.32 & 108.77
\\ 80 & 100.00 & 102.74 & 103.25 & 104.05 & 106.74
\\ 90 & 100.00 & 101.68 & 102.30 & 102.88 & 104.78
\\ 100 & 100.00 & 101.28 & 101.80 & 102.35 & 104.32
\end{tabular}
}\hfil
  \scalebox{0.85}{\begin{tabular}{r|ccccc}
  \multicolumn{6}{c}{{\bf MST+ (Improved stars)}} \\
  & Min & 1st q. & Median & 3rd q. & Max \\ \hline\hline
  0 & 101.51 & 108.48 & 109.72 & 110.91 & 115.01
\\ 10 & 101.51 & 108.15 & 109.34 & 110.74 & 114.01
\\ 20 & 101.51 & 107.61 & 108.91 & 110.09 & 115.50
\\ 30 & 101.01 & 106.92 & 107.99 & 109.42 & 114.56
\\ 40 & 100.06 & 106.00 & 107.28 & 108.45 & 112.86
\\ 50 & 100.10 & 105.24 & 106.41 & 107.40 & 111.18
\\ 60 & 100.10 & 104.27 & 105.36 & 106.48 & 110.55
\\ 70 & 100.00 & 103.44 & 104.21 & 105.21 & 108.77
\\ 80 & 100.00 & 102.64 & 103.16 & 103.89 & 106.19
\\ 90 & 100.00 & 101.71 & 102.28 & 102.86 & 104.67
\\ 100 & 100.00 & 101.21 & 101.78 & 102.22 & 104.32
\end{tabular}
} \medskip\\
  \scalebox{0.85}{\begin{tabular}{r|ccccc}
  \multicolumn{6}{c}{{\bf Zelikovsky}} \\
  & Min & 1st q. & Median & 3rd q. & Max \\ \hline\hline
  0 & 100.00 & 103.63 & 104.90 & 106.10 & 112.88
\\ 10 & 100.00 & 103.97 & 104.96 & 106.78 & 112.72
\\ 20 & 100.00 & 103.68 & 104.75 & 106.39 & 112.53
\\ 30 & 100.00 & 103.27 & 104.44 & 105.79 & 111.09
\\ 40 & 100.00 & 102.88 & 103.96 & 105.36 & 110.26
\\ 50 & 100.10 & 102.43 & 103.37 & 104.80 & 109.02
\\ 60 & 100.10 & 102.05 & 102.99 & 104.10 & 107.66
\\ 70 & 100.00 & 101.84 & 102.59 & 103.38 & 107.03
\\ 80 & 100.00 & 101.59 & 102.16 & 102.96 & 105.33
\\ 90 & 100.00 & 101.48 & 102.01 & 102.53 & 104.58
\\ 100 & 100.00 & 101.52 & 102.01 & 102.60 & 104.97
\end{tabular}
}\hfil
  \scalebox{0.85}{\begin{tabular}{r|ccccc}
  \multicolumn{6}{c}{{\bf Zelikovsky-}} \\
  & Min & 1st q. & Median & 3rd q. & Max \\ \hline\hline
  0 & 100.00 & 101.15 & 101.58 & 102.24 & 107.37
\\ 10 & 100.00 & 101.15 & 101.63 & 102.24 & 107.37
\\ 20 & 100.00 & 101.15 & 101.66 & 102.29 & 107.37
\\ 30 & 100.00 & 101.10 & 101.68 & 102.24 & 106.47
\\ 40 & 100.00 & 101.09 & 101.60 & 102.16 & 105.58
\\ 50 & 100.00 & 101.14 & 101.62 & 102.18 & 105.58
\\ 60 & 100.00 & 101.15 & 101.64 & 102.24 & 105.58
\\ 70 & 100.00 & 101.16 & 101.70 & 102.34 & 104.99
\\ 80 & 100.00 & 101.28 & 101.80 & 102.38 & 104.54
\\ 90 & 100.00 & 101.35 & 101.96 & 102.48 & 104.58
\\ 100 & 100.00 & 101.52 & 102.01 & 102.60 & 104.97
\end{tabular}
} \medskip\\
  \scalebox{0.85}{\begin{tabular}{r|ccccc}
  \multicolumn{6}{c}{{\bf Zelikovsky+}} \\
  & Min & 1st q. & Median & 3rd q. & Max \\ \hline\hline
  0 & 100.00 & 101.14 & 101.57 & 102.24 & 107.37
\\ 10 & 100.00 & 101.10 & 101.58 & 102.23 & 107.37
\\ 20 & 100.00 & 101.10 & 101.54 & 102.22 & 107.37
\\ 30 & 100.00 & 101.01 & 101.51 & 102.17 & 106.47
\\ 40 & 100.00 & 100.98 & 101.50 & 102.11 & 105.58
\\ 50 & 100.00 & 100.99 & 101.46 & 102.05 & 105.58
\\ 60 & 100.00 & 101.05 & 101.46 & 102.02 & 105.58
\\ 70 & 100.00 & 101.03 & 101.55 & 101.99 & 104.99
\\ 80 & 100.00 & 101.11 & 101.67 & 102.15 & 104.25
\\ 90 & 100.00 & 101.18 & 101.76 & 102.18 & 104.11
\\ 100 & 100.00 & 101.28 & 101.80 & 102.35 & 104.32
\end{tabular}
}
\medskip
  \caption{\label{tab:Rect}%
  Data used to generate Figure~\ref{fig:Rect} part I.}
\end{table*}

\FloatBarrier
\section{Figures on all instances but without Zelikovky's algorithm}\label{app:allInstances}

Due to time constraints original tests were perfomed only on 123 out of 200 instances from the PACE Challenge~2018.
The problem was running Zelikovsky's algorithm after each best star contraction which took too long on large instances.
Here, we provide figures where tests (excluding Zelikovsky's algorithm) were performed on almost all instances (excluding instances number 193, 196, 197, and 198 only)
Those four instances were still too large for testing.

\begin{figure*}[h!]
  \centering
  \includegraphics[width=0.49\textwidth]{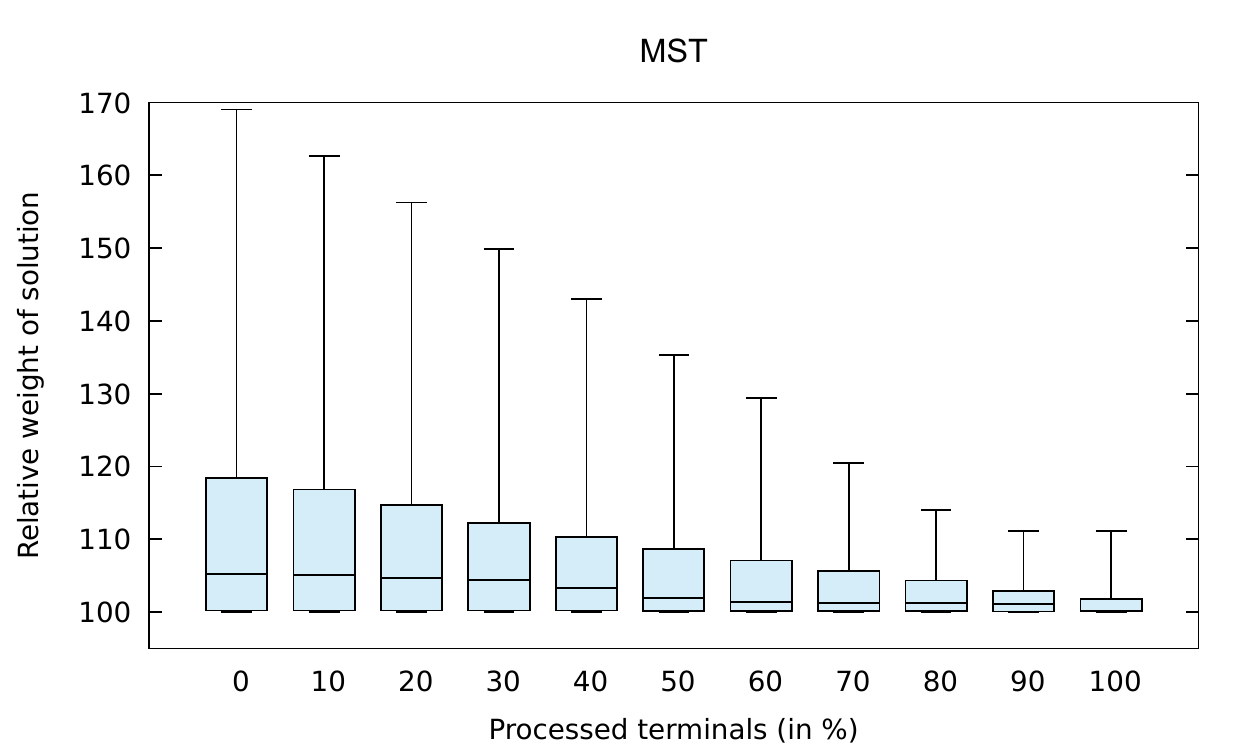}
  \includegraphics[width=0.49\textwidth]{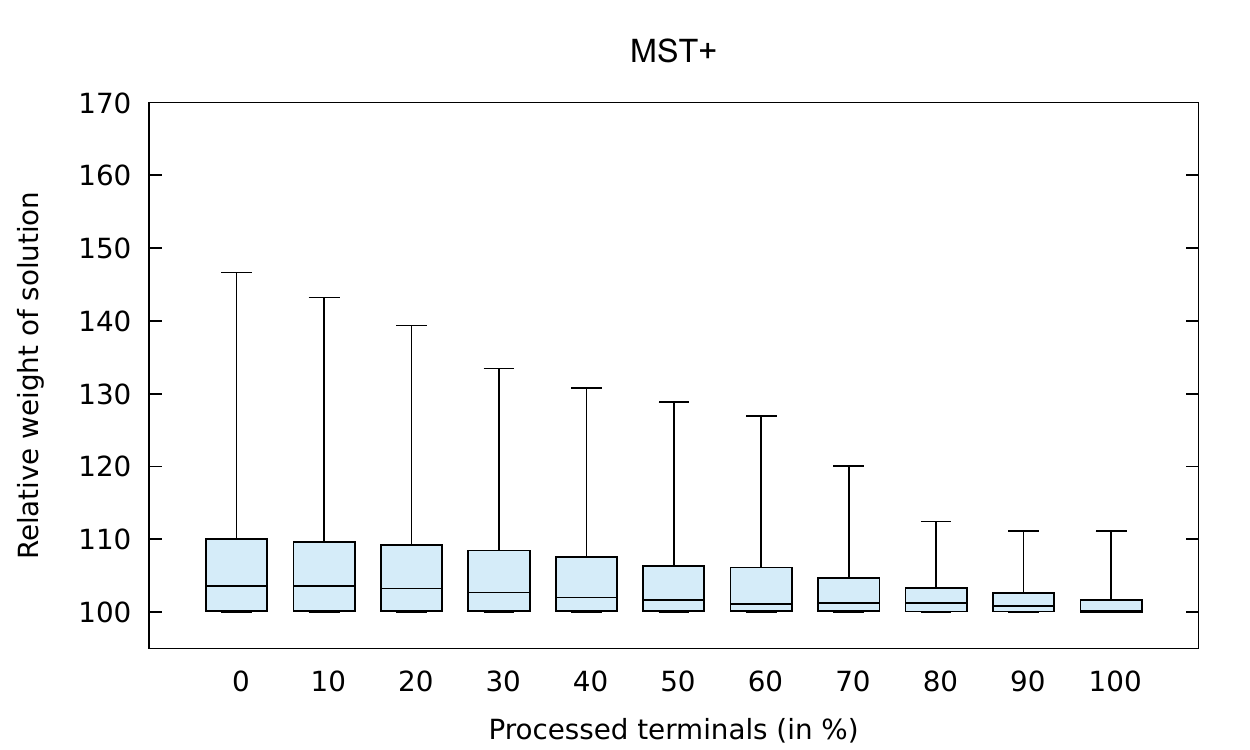}
  \includegraphics[width=0.49\textwidth]{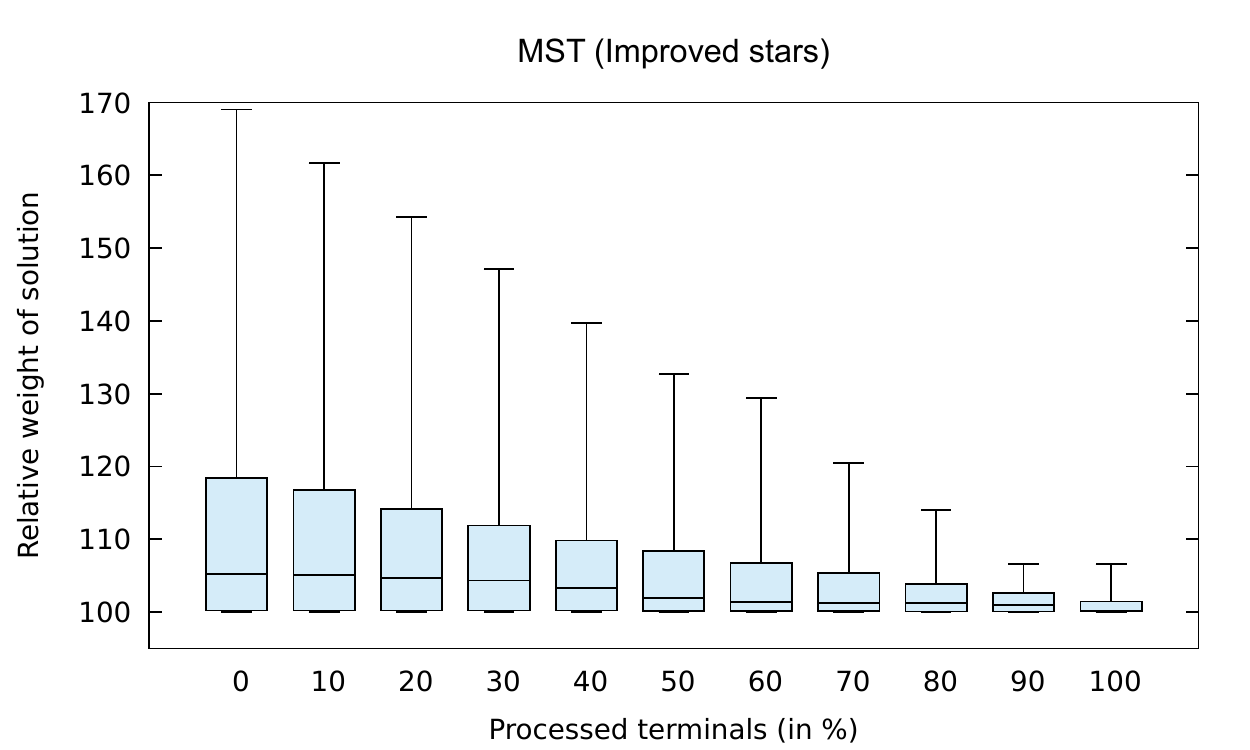}
  \includegraphics[width=0.49\textwidth]{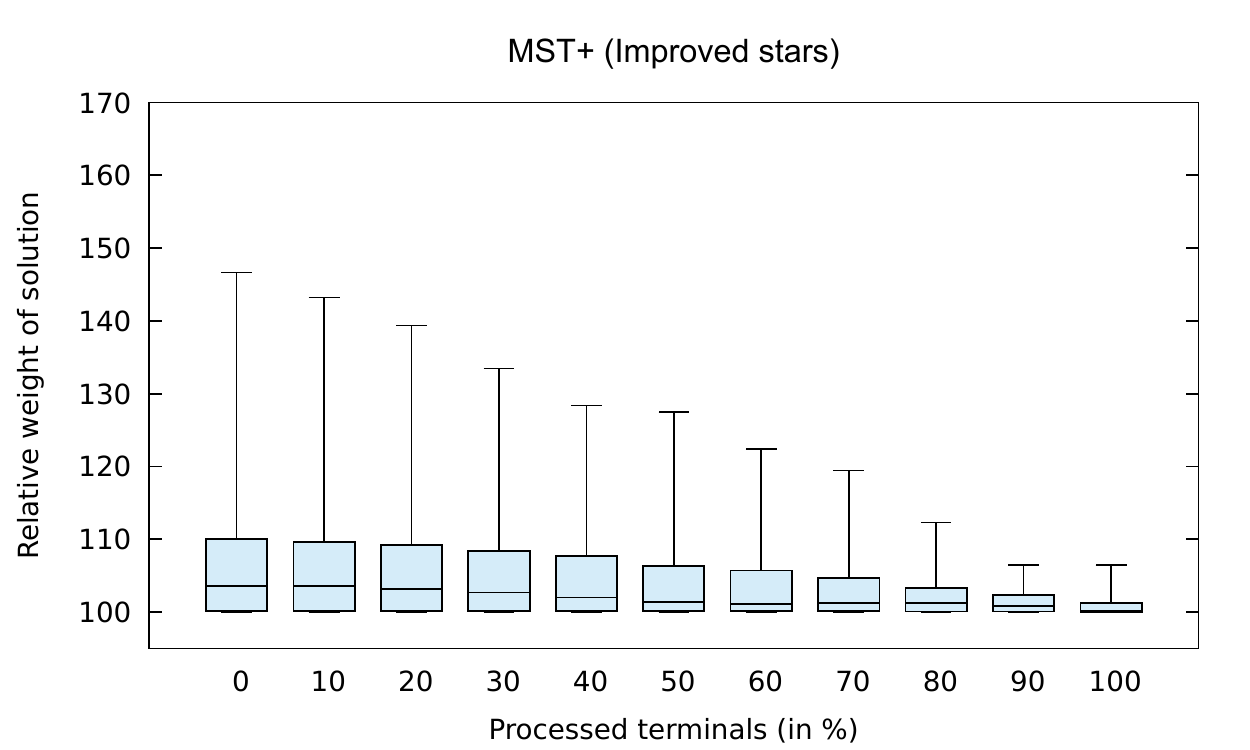}
  \includegraphics[width=0.49\textwidth]{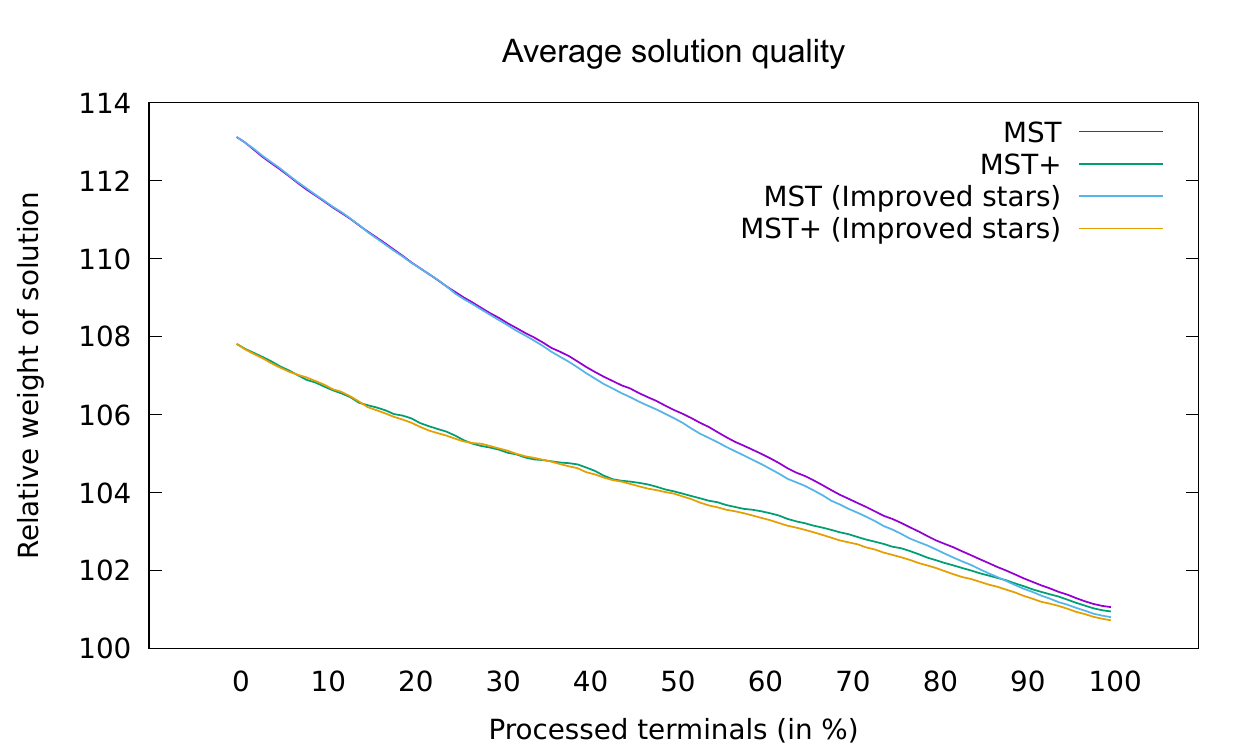}
\medskip
  \caption{%
    This chart shows the performance comparison of star contractions and MST heuristics
    on PACE Challenge 2018 instances.
}

\end{figure*}

\begin{figure*}[h!]
  \centering
  \includegraphics[width=0.48\textwidth]{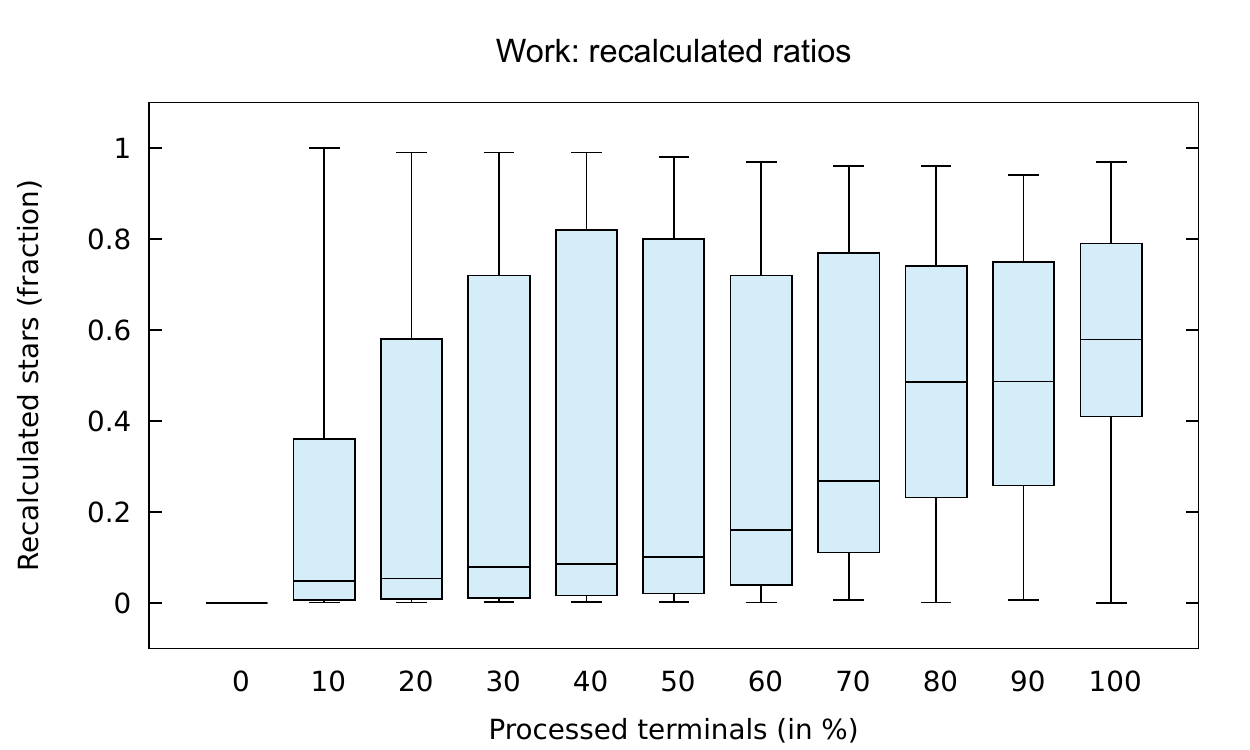}\hfil
  \includegraphics[width=0.48\textwidth]{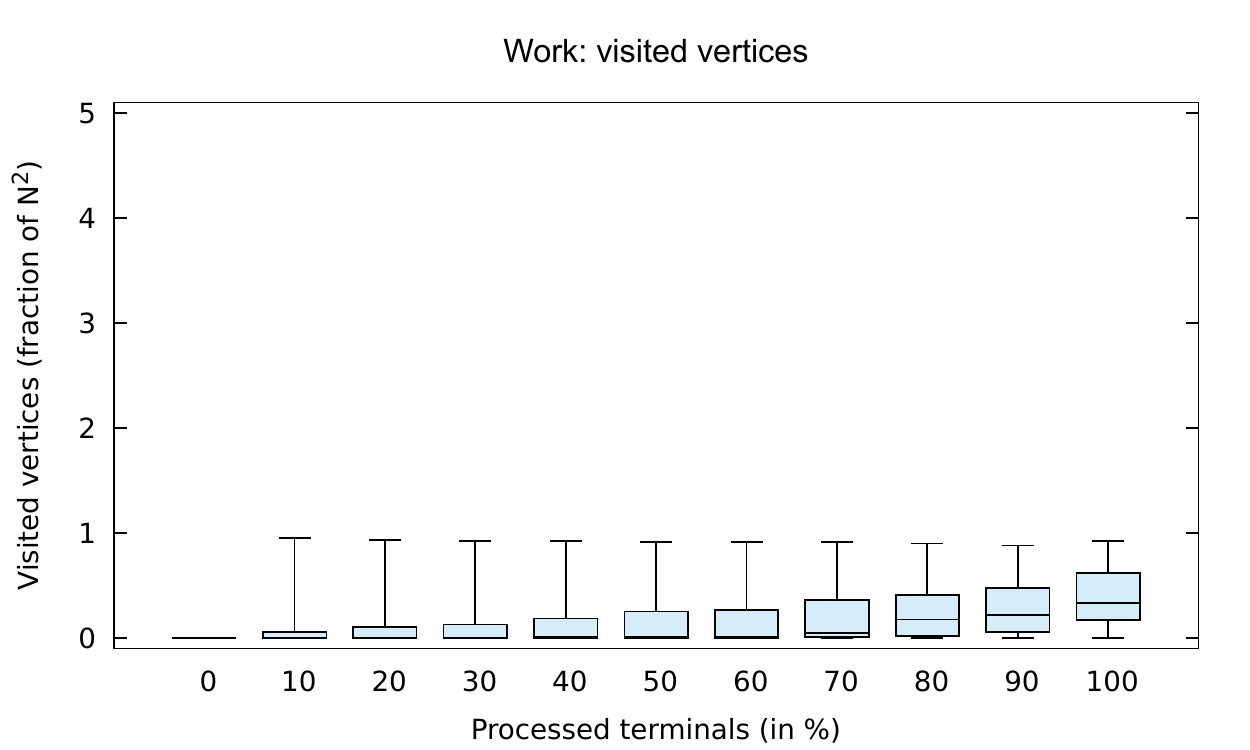}
  \includegraphics[width=0.48\textwidth]{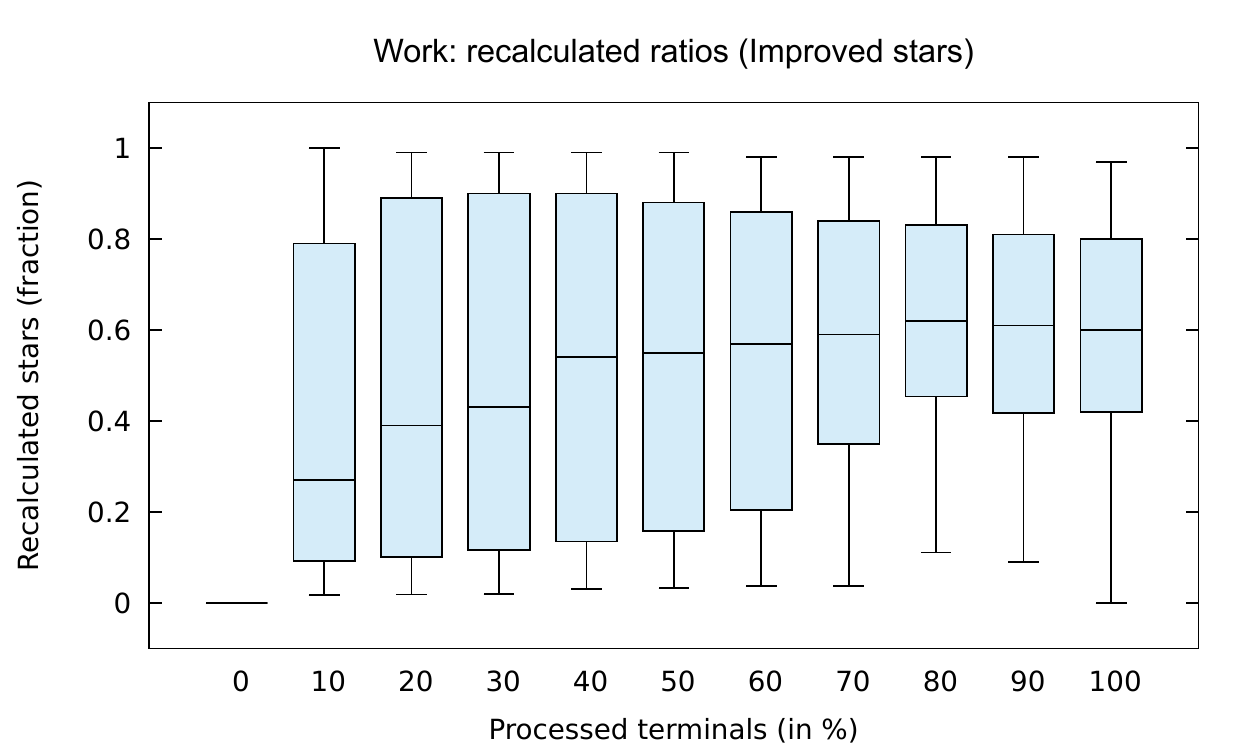}\hfil
  \includegraphics[width=0.48\textwidth]{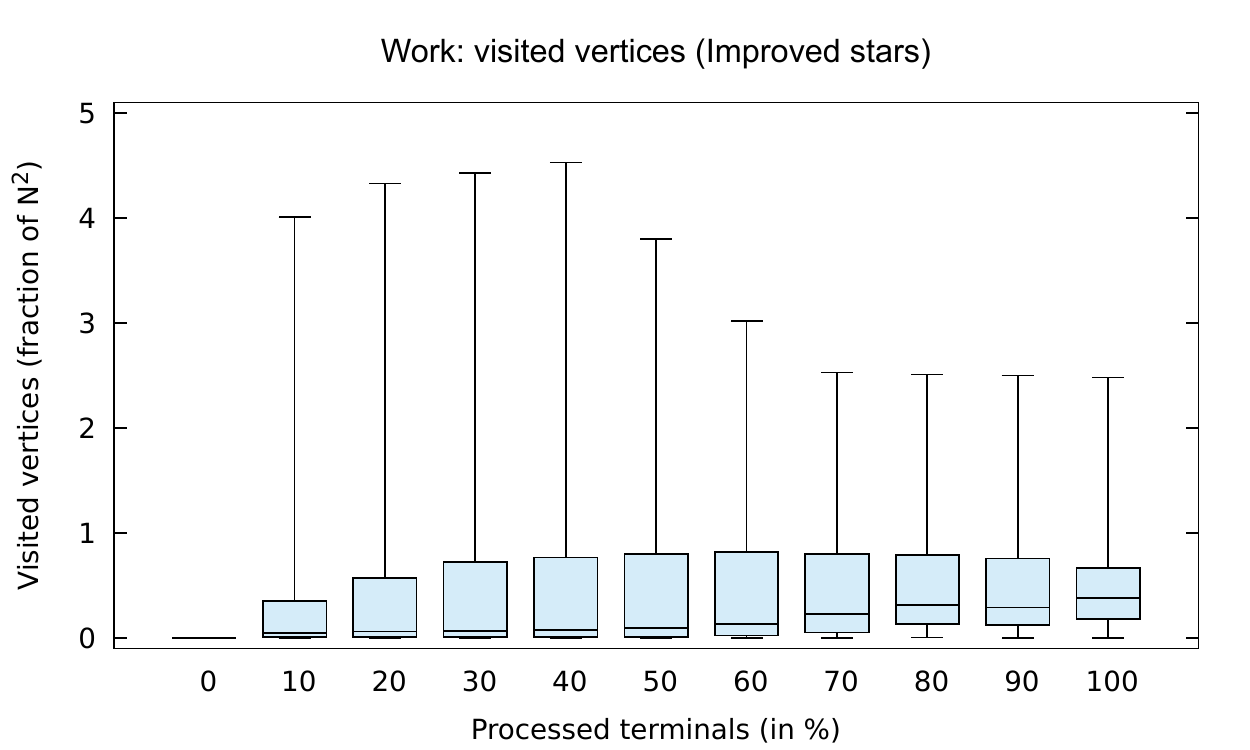}
  \includegraphics[width=0.48\textwidth]{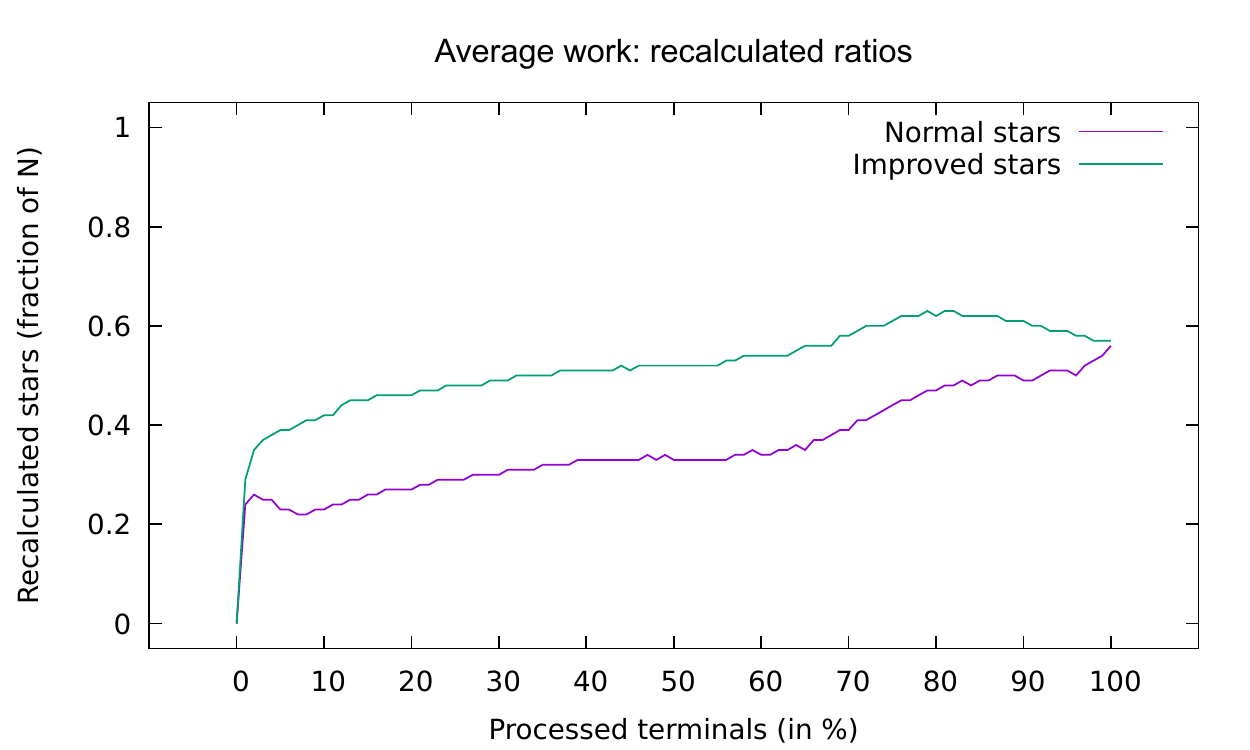}\hfil
  \includegraphics[width=0.48\textwidth]{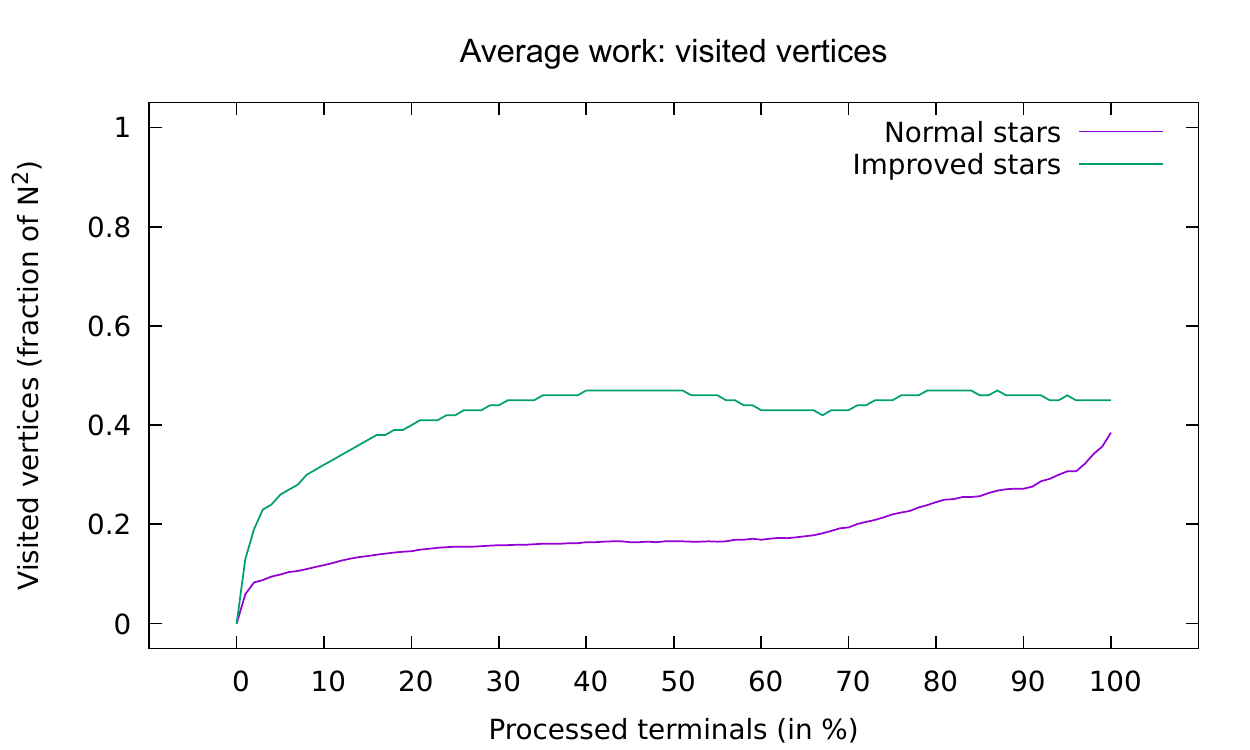}
\medskip
  \caption{%
  Work done on PACE Challenge 2018 instances.}
\end{figure*}

\begin{figure*}[h!]
  \begin{center}
  \includegraphics[width=0.8\textwidth]{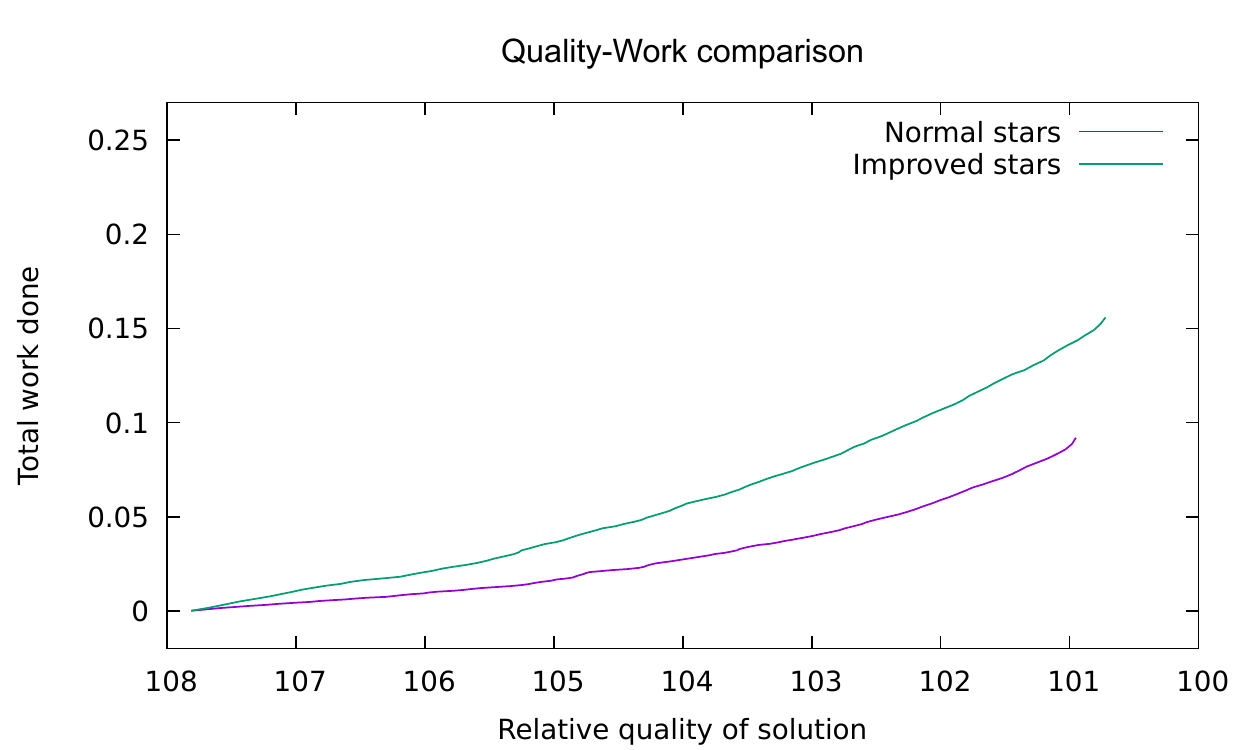}
\end{center}
\medskip
  \caption{%
  Work needed to get a solution of given quality using star contractions and MST+.
}
\end{figure*}

\end{document}